\documentclass[draft,nofootinbib,pre,twocolumn,showkeys,superscriptaddress,preprintnumbers,floatfix]{revtex4-1}

\usepackage{dynlearn}
\usepackage{physics}
\usepackage{mathrsfs}
\usepackage{comment}
\usepackage[english]{babel}
\usepackage{amssymb,amscd}
\newtheorem{theorem}{Theorem}

\newtheorem{definition}{Definition}

\newcommand{\MV}[1]{\left\langle #1 \right\rangle}

\newcommand{\PH}{\mathcal{P}(\mathcal{H})}

\newcommand{\CP}[1]{\mathbb{C}P^{#1}}
\newcommand{\QJK}[2]{Q( \vec{#1},\vec{#2})}
\newcommand{\DD}{\mathfrak{D}}

\newcommand{\iid}{i.i.d.\xspace}

\newtheorem{Definition}{Definition}

\def\tbf #1 {\textbf{#1} }

\begin{document}

\def\ourTitle{%
Quantum Information Dimension and Geometric Entropy
}

\def\ourAbstract{%
Geometric quantum mechanics, through its differential-geometric underpinning,
provides additional tools of analysis and interpretation that bring quantum
mechanics closer to classical mechanics: state spaces in both are continuous and 
equipped with symplectic geometry. This opens the door to revisiting foundational 
questions and issues, such as the nature of quantum entropy, from a geometric
perspective. Central to this is the concept of geometric quantum state---the
probability measure on a system's space of pure states. This space's continuity
leads us to introduce two analysis tools, inspired by Renyi's approach to information
theory of continuous variables, to characterize and quantify fundamental properties 
of geometric quantum states: the quantum information dimension, that is the rate of 
geometric quantum state compression, and the dimensional geometric entropy that 
monitors information stored in quantum states. We recount their classical definitions,
information-theoretic interpretations, and adapt them to quantum systems via the 
geometric approach. We then explicitly compute them in various examples and classes 
of quantum system. We conclude commenting on future directions for information in 
geometric quantum mechanics.
}

\def\ourKeywords{%
Quantum mechanics, geometric quantum mechanics, Renyi dimensions
}

\hypersetup{
  pdfauthor={Fabio Anza},
  pdftitle={\ourTitle},
  pdfsubject={\ourAbstract},
  pdfkeywords={\ourKeywords},
  pdfproducer={},
  pdfcreator={}
}


\title{\ourTitle}

\author{Fabio Anza}
\email{fanza@uw.edu}

\affiliation{InQubator for Quantum Simulations and Physics Department,\\
University of Washington, Seattle, WA 98195}

\affiliation{Complexity Sciences Center and Physics Department,
University of California at Davis, One Shields Avenue, Davis, CA 95616}

\author{James P. Crutchfield}
\email{chaos@ucdavis.edu}

\affiliation{Complexity Sciences Center and Physics Department,
University of California at Davis, One Shields Avenue, Davis, CA 95616}

\date{\today}
\bibliographystyle{unsrt}

\begin{abstract}
\ourAbstract
\end{abstract}



\preprint{\arxiv{2111.06374}}

\date{\today}
\maketitle





\section{Introduction}
\label{sec:Intro}

When connecting theory to experiment both classical and quantum mechanics (CM
and QM) must cope with the emergence of randomness and uncertainty. However,
the nature of randomness, and its dynamical emergence, can differ. Building on
previous results \cite{Anza20a,Sone21} that exploit geometric parallels between
classical and quantum state spaces (both with state spaces equipped with
symplectic manifolds), we extend several tools for analyzing out-of-equilibrium
classical systems to the quantum domain. This strengthens the parallels and
provides a novel paradigm for investigating far-from-equilibrium open quantum
systems.

Specifically, following Kolmogorov and Sinai's use of Shannon's information
theory \cite{Shan48a} to quantify degrees of deterministic chaos
\cite{Gelf56a,Kolm56a,Kolm59b,Kolm58,Kolm59,Sina59}, we show that the parallels
go even deeper and lead to new descriptive and quantitative tools. This is done
using \emph{Geometric Quantum Mechanics} (QGM), an approach to quantum
mechanics based on differential geometry that removes physical redundancies in
quantum states intrinsic to the standard, linear-algebra approach.

QM, in point of fact, is grounded in a formalism in which the states of a
discrete system are vectors in a complex Hilbert space $\mathcal{H}$ of generic
finite dimension $D$. However, it is well-known that such a formulation is
redundant: vectors differing only in normalization and global phase are
physically equivalent. Implementing this equivalence relation leads to the
space where quantum states live: the complex projective Hilbert space $\PH \sim
\mathbb{C}P^{D-1}$. This is GQM's starting point 
\cite{STROCCHI1966,Miel68,Kibble1979,Heslot1985,Page87,And90,Gibbons1992,Ashtekar1995,Ashtekar1999,Brody2001,Carinena2007,Chruscinski2006,Marmo2010,Avron2020,Pastorello2015,Pastorello2015a,Pastorello2016,Clemente-Gallardo2013}.

It is important to stress that, while the mathematical formulation differs, the
phenomena addressed are precisely the same as standard quantum mechanics.
Nonetheless, the geometric approach has proven (i) to be a rich source of
fundamental insights into the nature of quantum phenomena and (ii) to lead to
powerful analysis tools. Our goal is to advance this perspective to investigate
the out-of-equilibrium phenomenology of open quantum systems.

GQM works with probability measures on $\PH$. These are interpreted using
ensemble theory, as noninteracting copies of pure states for the same quantum
system, distributed according to some measure. This leads to the concept of a
\emph{geometric quantum state} (GQS) as an ensemble of pure states. (For an
extensive analysis we recommend Ref. \cite{Brody2001}.) This is a more
fundamental notion of quantum state than the density matrix, as the latter can
be computed from the former, but not vice versa. 

Recent work provided a constructive procedure to compute the GQS of an open,
finite-dimensional, quantum system interacting with another one of arbitrary
(finite or infinite) dimension \cite{Anza20a}. This revealed why the GQS
provides a more accurate description than available with a density matrix: The
GQS retains the details about how a specific ensemble of pure states emerges
from the structure of correlations between the system and its surroundings. 

Starting from this foundation, the following introduces two
information-theoretic concepts to characterize geometric quantum states. The
first is the \emph{quantum information dimension}. This borrows from Renyi's notion of
the effective dimension of a continuous probability distribution, developed in
the setting of efficiently transmitting continuous variables over noisy
communication channels. Interestingly, for classical variables there are
distributions for which the dimension is not an integer---these are the well-known 
\emph{fractals} \cite{Feld12}. It also has an
operational interpretation within communication theory, as the upper bound
on the lossless compression rate for transmitting GQSs. The second GQS
characterization uses the (related) concept of dimensional geometric entropy.
Accounting for GQS dimension, this entropy quantifies the information a GQS
stores about a quantum system.

The development unfolds as follows. Section \ref{sec:GQM} gives a brief summary
of geometric quantum mechanics and the notion of geometric quantum state.
Section \ref{sec:QID} defines the quantum information dimension, while Section
\ref{sec:GDQE} introduces the dimensional geometric entropy. Sections
\ref{sec:Example1} to \ref{sec:Example4} then analyze several examples,
evaluating these quantities exactly. The first is an open quantum system
interacting with a finite-dimensional environment. The second is an open
quantum system interacting with another with an infinite-dimensional Hilbert
space (continuous degrees of freedom). The third shows how to evaluate these
quantities for discrete-time chaotic quantum dynamics. The fourth shows how to
evaluate these quantities in the thermodynamic limit for a condensed-matter
system in its ground state. Finally, Section \ref{sec:Discussion} discusses the
results and Section \ref{sec:Conclusions} draws forward-looking conclusions.
 
\section{Geometric Quantum Mechanics}
\label{sec:GQM}

References
\cite{STROCCHI1966,Miel68,Kibble1979,Heslot1985,Page87,And90,Gibbons1992,Ashtekar1995,Ashtekar1999,Brody2001,Carinena2007,Chruscinski2006,Marmo2010,Avron2020,Pastorello2015,Pastorello2015a,Pastorello2016,Clemente-Gallardo2013}
lay out the mathematical physics of GQM. Here, we simply
recall the aspects most relevant for our purposes. Throughout, we only address
quantum systems with a Hilbert space $\mathcal{H}$ of finite dimension $D$. In
GQM, the pure states of such systems are points $Z$ in the complex projective
space $\PH \sim \mathbb{C}P^{D-1}$. Given an arbitrary basis
$\left\{\ket{b_{\alpha}}\right\}_{\alpha}$ of $\mathcal{H}$, the pure state
$Z$ has the vector representation:
\begin{align*}
\ket{Z} \coloneqq \sum_{\alpha=0}^{D-1} Z_\alpha \ket{b_\alpha}\in \mathcal{H}
  ~,
\end{align*}
where $Z_\alpha \in \mathbb{C}$ and $Z \sim \lambda Z$, with $\lambda \in
\mathbb{C}$, giving $Z\in \mathbb{C}P^{D-1}$. This space has a rich geometric structure \cite{Bengtsson2017}. In
particular, there is a well-defined metric---the \emph{Fubini-Study metric}
$g_{FS}$---and a related notion of Volume $dV_{FS}$---the Fubini-Study volume
element. These are directly connected, up to an overall positive multiplicative scalar, by 
$dV_{FS} = \sqrt{\det g_{FS}}dZ d\overline{Z}$, where the overbar is the complex conjugate 
and $dZ$ is Lebesgue measure. While a full explication is beyond our current scope, 
we simply give its form in a particular coordinate system, specified by $Z_\alpha =
\sqrt{p_\alpha}e^{i\phi_\alpha}$: $dV_{FS} = \prod_{\alpha=1}^{D-1}
\left(dp_\alpha d\phi_\alpha / 2\right)$. 

On $\PH$, one considers ensembles distributed according to a probability
density function or, more generally, a probability measure $\mu$. The simplest example is
the basic definition of the uniform measure: $d\nu_{FS} \coloneqq dV_{FS} /
V_{D-1}$, where the total Fubini-Study volume of $\PH$ is $V_{D-1}
= \pi^{D-1} / (D-1)!$. This determines the basic notion of uniform measure
on $\PH$. Calling $A$ an element of $\PH$'s Borel $\sigma$-algebra and adopting
the De Finetti notation, we have:
\begin{align*}
\nu_{FS}(A) := \frac{1}{V_{D-1}} \int_{A} dV_{FS}
  ~.
\end{align*}

In general cases, the measure $\mu$ is not uniform and one has:
\begin{align*}
\mu(A) = \int_{A} \mu(d\nu_{FS})
  ~.
\end{align*}
Looking at the measure-theoretic definition, if $\mu$ is absolutely continuous
with respect to $\nu_{FS}$, then there is a probability density function $q(Z =
z)$ such that:
\begin{align*}
\mu_q(A) = \int_A q(z) d\nu_{FS}
  ~.
\end{align*}
This is interpreted by saying that $\mu_q(d\nu_{FS}^z) = q(z)d\nu_{FS}^z$ is
the infinitesimal probability of a realization $z$---i.e., a system pure
state---belonging to an infinitesimal volume $d\nu_{FS}^z$ centered at $z \in
\PH$. Thus, we can think about the pure state of a quantum system as a
realization $z$ of a random variable $Z$ with sample space $\PH$. Here, $z$ is
distributed on $\PH$ according to its geometric quantum state $q(z)$ or $\mu$.
Through an abuse of language, we often refer to both the measure $\mu$ and the
density $q$ (when it exists) as geometric quantum states. This is acceptable as
they convey the same kind of information. 

\begin{figure}[b]
\centering
\includegraphics[width=.9\columnwidth]{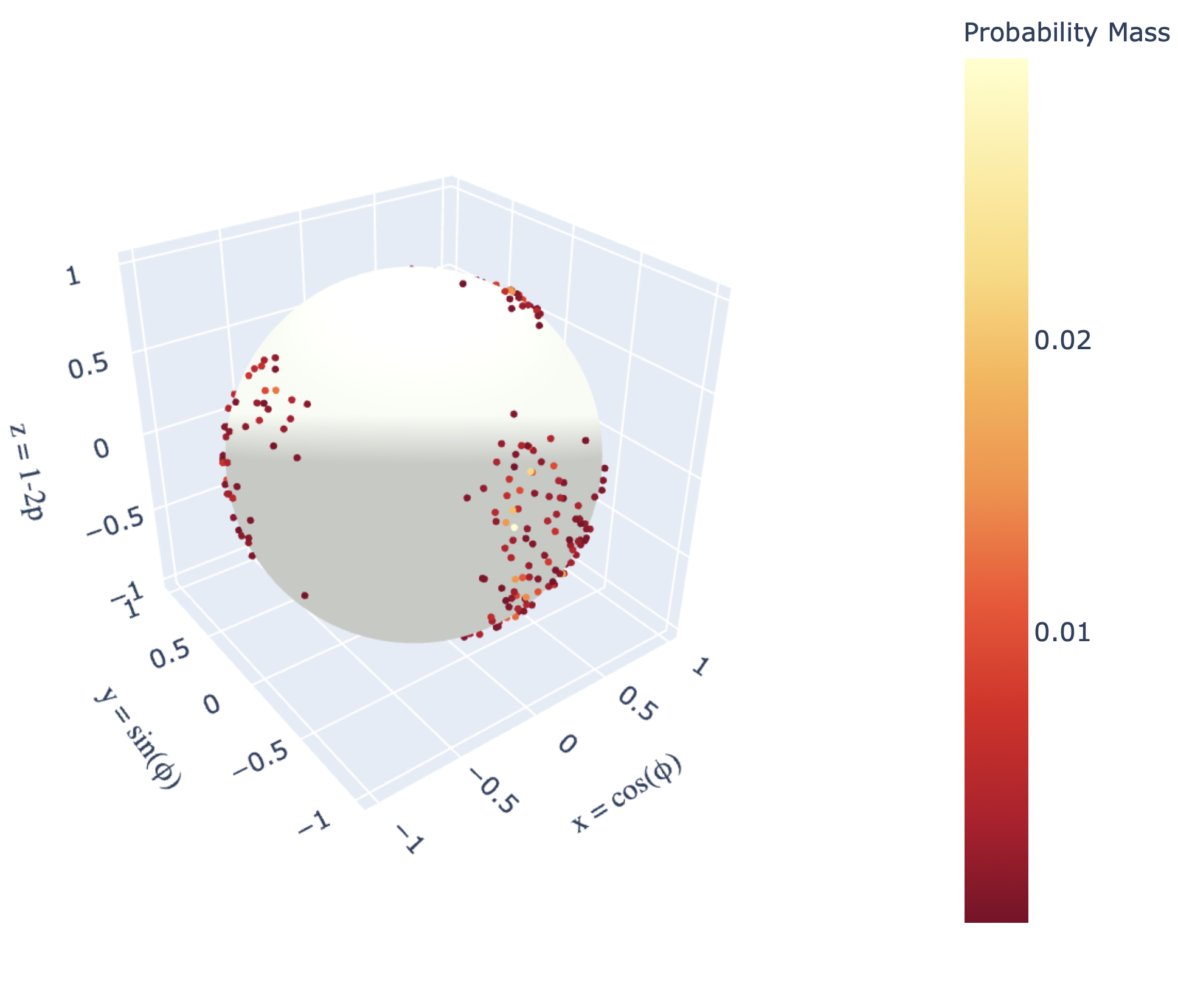}
\caption{Geometric quantum state $q(Z) = \sum_{k=1}^{2^9} p_k\delta[Z-Z_k]$
	on $\mathbb{C}P^1$, with coordinates $(\theta,\phi)$. It is a finite sum
	of $2^9$ Dirac measures. Each point is a possible pure state
	$\ket{Z_k} = \cos \theta_k/2 \ket{0} + \sin \theta_k/2
	e^{i\phi_k}\ket{1}$ in which the system can be, with probability $p_k$. The
	specific value of each $p_k$ is encoded in a point's color; see legend.
	}
\label{fig:GQS} 
\end{figure}

Together, the triple $(\PH,Z,\mu)$ defines a random variable, in the classical
sense, in which the sample space is continuous and encodes the underlying
quantumness of the physical system we aim to describe. We call this a
\emph{random quantum variable} (RQV). Following standard notation, if $Z$ is
an RQV, we denote a realization with a lowercase corresponding letter $z$.
(This is not to be confused with the notation $Z_\alpha$, with a Greek label,
that refers to the $\alpha$-th component of the vector $\ket{Z} \in \mathcal{H}$.

To aid intuition, Fig. \ref{fig:GQS} displays an example of a geometric quantum
state of a qubit, using the $(\theta,\phi)$ coordinates to represent $\PH$ as
the familiar Bloch sphere.

\section{Quantum Information Dimension}
\label{sec:QID}

One GQM advantage we exploit is that it allows the use of classical measure
theory. The price paid is working with an underlying manifold of nontrivial geometry, 
as just noted. Intuitively, GQM directly encodes ``quantumness'' in the underlying 
geometry of the sampling space. Building on this, we now formalize the dimension 
of a geometric quantum state, extending Renyi's information theory \cite{Renyi59} to 
the quantum domain.

Given $\PH$ and a measure $\mu$ on it, we uniformly discretize the manifold
and coarse-grain $\mu$ to obtain a discrete distribution. This is accomplished 
as follows, using ``probabilities + phases'' coordinates: 
$Z_\alpha = \sqrt{p_\alpha}e^{i\phi_\alpha}$. In this,
$\left\{p_\alpha\right\}_{\alpha=0}^{D-1}$
lives in a  probability simplex $\Delta_{D}$, while the phase coordinates 
$\left\{\phi_\alpha\right\}_{\alpha=1}^{D-1}$ live on a $(D-1)$-dimensional 
torus $\mathbb{T}^{D-1}$.

Given this coordinate set, we partition both $\Delta_D$ and
$\mathbb{T}_{D-1}$ separately, in a uniform fashion. More accurately,
this is a partitioning of the $\sigma$-algebra on $\PH$ into a finite and
discrete collection of sets. Since both $\Delta_D$ and $\mathbb{T}_{D-1}$ are
manifolds of real dimension $D-1$, this generates a uniform partition $\left\{
Q(\vec{j},\vec{k})\right\}_{\vec{j},\vec{k}}$ of $\PH$ with a total number of
$L^{D-1} \times L^{D-1}$ cells. This means $\bigcup_{j_1,\ldots,j_{D-1}=1}^{L}
\bigcup_{k_1,\ldots,k_{D-1}=1}^{L} Q(\vec{j},\vec{k}) = \PH$, with
$Q(\vec{j},\vec{k}) \cap Q(\vec{i},\vec{m}) =
\delta_{\vec{j},\vec{i}}\delta_{\vec{k},\vec{m}}Q(\vec{i},\vec{m})$. Here,
$\vec{j} = (j_1,\ldots,j_{D-1})$ is the multi-index label that runs over the
discretization $\left\{ \Delta_D(\vec{j})\right\}_{\vec{j}}$ of $\Delta_D$,
with $j_\alpha = 1,\ldots, L$ and $\vec{k} = (k_1,\ldots, k_{D-1})$ is the
multi-index label that runs over the discretization $\left\{
\mathbb{T}_{D-1}(\vec{k})\right\}_{\vec{k}}$ of $\mathbb{T}_{D-1}$, with
$k_\alpha = 1,\ldots,L$.

The reason to partition $\PH$ using this coordinate system is the resulting
factorization of $d\nu_{FS}$ in ``$d p \times d \phi$'':
\begin{align*}
d\nu_{FS} = dP(\Delta_D) dQ(\mathbb{T}_{D-1})
\end{align*}
where $dP(\Delta_D) = (D-1)! \delta \left(\sum_{\alpha=0}^{D-1}p_\alpha - 1
\right) dp_0 \ldots dp_{D-1}$ is the flat measure on $\Delta_D$ and
$dQ(\mathbb{T}_{D-1})$ is the flat measure on $\mathbb{T}_{D-1}$.

This allows us to separately discretize the probability simplex and the
high-dimensional torus, while also resulting in cells $Q(\vec{j},\vec{k})$ 
with uniform Fubini-Study volume:
\begin{align}
V_{D-1}(\epsilon) & \coloneqq \nu_{FS}(Q(\vec{j},\vec{k}))
  \nonumber \\
  & = \frac{1}{L^{2(D-1)}} \nonumber \\
  & = \epsilon^{-2(D-1)}
  ~,
\label{eq:volume}
\end{align}
where $\epsilon$ sets the scale of the partition elements. Note that, despite
the discretization's specific coordinate system, the value of
$V_{D-1}(\epsilon)$ does not depend on the coordinate system, thanks to the 
invariance of $d\nu_{FS}$. Therefore, $\left\{
Q(\vec{j},\vec{k})\right\}_{\vec{j},\vec{k}}$ defines a good uniform partition
of $\PH$.

Calling $z_{\vec{j},\vec{k}}$ a generic point in $Q(\vec{j},\vec{k})$, we can
now perform the coarse-graining procedure to get a discrete random variable
$Z^\epsilon$, for $Z \in \CP{D-1}$:
\begin{align*}
Z \in Q(\vec{j},\vec{k}) \quad \Rightarrow \quad Z^\epsilon = z_{\vec{j},\vec{k}}
  ~,
\end{align*}
where the probability $p_{\vec{j},\vec{k}}$ of $Z^\epsilon \in \QJK{j}{k}$ is
defined via coarse-graining:
\begin{align}
\mathrm{Pr} \left( Z \in \QJK{j}{k}\right) &:= p_{\vec{j},\vec{k}} \nonumber \\
  & = \mu(\QJK{j}{k}) \nonumber \\ 
  & = q(z_{\vec{j},\vec{k}}) V_{D-1}(\epsilon) 
  ~,
\label{eq:coarse}
\end{align}
where (only) in the last equality we assumed the existence of a density $q(Z)$
and, in that case, $z_{\vec{j},\vec{k}}$ is a point inside $Q(\vec{j},\vec{k})$
defined implicitly by $q(z_{\vec{j},\vec{k}}) = \mu(Q(\vec{j},\vec{k})) /
V_{D-1}(\epsilon)$.

For explorations in discretizing continuous random variables see the lectures
in Ref. \cite{Graf00} and relevant developments in Refs. \cite{Brig19,Pag15}.

Given that $Z^\epsilon$ is a discrete random variable, we can now use Shannon's
functional to evaluate its entropy:
\begin{align}
H\left( Z^\epsilon \right)
  = - \sum_{\vec{j},\vec{k}} p_{\vec{j},\vec{k}} \log p_{\vec{j},\vec{k}}
  ~.
\label{eq:DiscretizedShannon}
\end{align}
Substituting Eq. (\ref{eq:volume}) into Eq. (\ref{eq:coarse}) and calling
$\mathfrak{D}$ the \emph{real dimension} of the submanifold of $\PH$ on which
$\mu$ has nonzero support, then as $\epsilon \to 0$ one has that
$H(Z^\epsilon)$ is linear in $-\log \epsilon$ \cite{Renyi59}:
\begin{align*}
H(Z^\epsilon) \sim_{\epsilon \to 0} - \mathfrak{D} \log \epsilon + \mathfrak{h}
  ~.
\end{align*}
Note that $\mu$ can also have support on the whole manifold, in which case
$\mathfrak{D} = 2(D-1)$---the real dimension of $\PH$. Thus, $\mathfrak{D}$'s
value in this scaling is the \emph{information dimension} while, as we show
shortly, the offset $\mathfrak{H}$ provides a workable definition of
differential entropy that accounts for the dimension of $\mu$'s support.

With this in mind, we are now ready to define a geometric quantum state's
\emph{quantum information dimension}.

\begin{Definition}[Quantum Information Dimension]
Given a finite-dimensional quantum system with state space $\PH \sim \CP{D-1}$ and 
geometric quantum state $\mu$, the latter's quantum information dimension $\mathfrak{D}$ is:
\begin{align}
\mathfrak{D}
  \coloneqq \lim _{\epsilon \to 0} \frac{H\left( Z^\epsilon\right)}{- \log \epsilon}\label{eq:D}
  ~.
\end{align}
\end{Definition}

While alternative definitions are possible, see Ref. \cite{Wu10} and references
therein, they all aim to rigorously ground the same idea. And, the
alternatives often provide identical values, assuming the validity of some
regularity conditions. The key point is that the essence and the result of the
development do not change under these alternatives. For a detailed explication
of information dimension see Refs. \cite{Renyi59,Wu10,Farm83}.

Note that there are useful theorems for explicitly calculating the information
dimension. We use these, and develop some other ones, in the upcoming sections.
Before proceeding to the geometric dimensional quantum entropy, though, we
briefly discuss a connection between $\mathfrak{D}$ and analog information
theory, where $\mathfrak{D}$'s classical counterpart has a direct
interpretation.

\section{$\mathfrak{D}$'s Information-theoretic Underpinning}

Quantum information theory takes inspiration from the information theory of
classical discrete sources. However, it is well known that quantum states need
real numbers to be faithfully represented. In fact, they require several complex
numbers or, equivalently, elements of $\mathbb{R}^{2n}$. So, an approach inspired
by information theory is appropriate \cite{Cove91a} if we can identify a
natural extension to situations where the random variables at hand have a
continuous sample space. As such, one can also appeal to analog or continuous
information theory. An example, relevant for our purposes, of a result from
analog information theory is the quasi-lossless compression theorem. Loosely
speaking, this answers the question ``How much can we compress the information
emitted by a continuous source, using continuous variables?''

Rather than giving the full result---for which see Ref. \cite{Wu10}---we simply
discuss the essential point. Consider a continuous source emitting realizations
$\vec{x} \in \mathcal{X}$ of a random variable $X$. We desire to compress its
information. The dimension of $\mathcal{X}$ is arbitrary, but we assume
$\mathcal{X} \subseteq \mathbb{R}^n$, for some $n$.

Compression can be achieved using $(N,K)$-codes---a pair of encoder-decoder
functions. The encoder function $f : \mathcal{X}^N \to \mathcal{Y}^K$ converts
the continuous message into appropriate discrete symbols, belonging to the
space $\mathcal{Y}$. The decoder function $g : \mathcal{Y}^K \to
\mathcal{X}^N$ performs the inverse. 

Take the probability of making an error as $\delta = \mathrm{Pr}\left[
g(f(\vec{x})) \neq \vec{x}\right]$. Call $R(\epsilon)$ the infimum of $R \geq
0$ such that the $\left( N ,\lfloor RN\rfloor \right)$ code has $\delta \leq
\epsilon$ error. Assuming a linear form for the encoder and decoder, one
establishes that there is a fundamental limit to the amount of quasi-lossless
(up to $\epsilon$) compression one can reach. This limit is achievable and it
is given by the source's classical information dimension: $R(\epsilon)
\leq \mathfrak{D}(X)$.

Here, with a slight abuse of notation, we use the same symbol $\mathfrak{D}$ to
also identify the classical information dimension. We also stress that this is
only a brief and simplified summary of the comprehensive analysis performed in
Ref. \cite{Wu10}. What is relevant for our purposes is the fact that, despite
its simplicity, it is directly applicable to quantum systems. In particular, it
addresses encoding a quantum source emitting pure states $Z \in \PH$ with
a classical continuous distribution given by the geometric quantum state $\mu$.

Moreover, as quantum states themselves are points on a manifold described by
continuous variables, it can also be applied to the inverse problem of
representing a continuous classical source with quantum states. While this begs
further exploration before making rigorous statements, we believe it hints at
the fact that there is an alternative way, inspired by analog information
theory, of conceptualizing quantum computing and information theory.

Before finally moving to dimensional quantum entropy, we highlight a point about
$\mathfrak{D}$. While the understanding based on encoding and communication
theory strengthens the argument for relevance, $\mathfrak{D}$'s general role in
investigating properties of geometric quantum states stands on its own, as it is
independently and rigorously defined.

\section{Dimensional Quantum Entropy}
\label{sec:GDQE}

For a given geometric quantum state, $\DD$ gives a notion of effective
dimension. It is therefore natural that its value affects the definition of
entropy one assigns to a geometric quantum state.

The standard example comes from comparing discrete and continuous probability
distributions. In the discrete setting there is a unique entropy definition,
given by Shannon's functional:
\begin{align}
H_{\mathrm{discrete}} = - \sum_i p_i \log p_i
  ~.
\label{eq:DiscreteEntropy}
\end{align}
Its extension to the continuous domain, however, is not unique and its
construction, use, and interpretation require care. On the one hand, Shannon's
original definition of differential entropy for a continuous variable $X$ with
probability distribution $p(x)$ provides a meaningful and interpretable
quantity \cite{Shan48a}:
\begin{align}
H_{\mathrm{continuous}} = - \int p(x) \log p(x) dx
  ~.
\label{eq:DiffEntropy}
\end{align}
On the other, it is well known that it presents its own challenges and that
alternatives are possible. For example, it is well-known to be sensitive to
rescaling of the measure. When $dx \to k dx$ we have that
$H_{\mathrm{continuous}} \to H_{\mathrm{continuous}} + \log k$.

Thus, when a physical measure is defined up to an overall scale factor, this
quantity is defined up to an overall additive factor. This is an issue that can
often be disregarded as it does not carry physical consequences, in analogy
with the classical notion of energy, defined up to a constant. Practically,
this can be bypassed by fixing the zero point of the entropy to be given by the
uniform distribution. This is realized by taking the measure to be the
normalized volume of the space. In this way, a uniform density simply has
constant value equal to $1$, giving a differential entropy $H_{unif} = \log 1
=0$.

Note, too, that $H_{\mathrm{continuous}}$ can be negative, as $-\log p(x)$ can
be negative when $p(x)$ is a density. This is not a concern, since correctly
interpreting this quantity relies on the asymptotic equipartition property,
which holds for both discrete and continuous random variables, irrespective of
$H_{\mathrm{continuous}}$'s sign; see Ch. 8 of Ref. \cite{Cove91a}.

Moreover and finally, the differential entropy is appropriate only when the
distribution has integer topological dimension. This is not true, for example,
in nonlinear dynamics, in which time-asymptotic statistical states often live
on fractals due, for example, to chaotic behavior. These objects do not have
integer dimension. However, it is possible to define an entropy that takes this
rich phenomenology into account. Again, for the classical result we point to
Refs. \cite{Renyi59,Wu10}. Here, we extend this into the quantum domain as
follows.

\begin{Definition}[Dimensional quantum entropy]
Given a finite-dimensional quantum system with state space $\CP{D-1}$, geometric
quantum state $\mu$ with quantum information dimension $\mathfrak{D}$, we define
$\mu$'s dimensional quantum entropy $H_{\mathfrak{D}}\left[\mu\right]$ as:
\begin{align}
H_{\mathfrak{D}}\left[\mu\right] \coloneqq \lim_{\epsilon \to 0}
	\big( H\left(Z^\epsilon\right) + \mathfrak{D} \log \epsilon\big)
  ~.
\label{eq:GDQE}
\end{align}
\end{Definition}

Note that this entropy is parametrized by the quantum information dimension. To
provide intuition, consider two simple cases. Shortly after, Secs. \ref{sec:Example1} 
to \ref{sec:Example4} present a series of examples, with detailed calculations.

First, if $\DD = 0$, we see that $H_{0}\left[\mu\right]$ is simply the
continuum limit of $Z^\epsilon$'s entropy $H(Z^\epsilon)$. Second, imagine we
are looking at the uniform distribution over Bloch sphere $\CP{1}$. As this is
an absolutely continuous distribution, it has quantum information dimension
$\DD = 2$ and, therefore, the appropriate notion of entropy should take that
into account.

We also find that when $\DD=2(D-1)$, $H_{\DD=2(D-1)}[Z]$ is equal to the notion
of geometric quantum entropy introduced, as far as we know, by Ref.
\cite{Brody2000}. See also Refs.
\cite{Brody1998,Brody2000,Brody2001,Brody2007,Brody2001,Anza20a,Anza20b,Anza20c,Brody2016}.
In the simple case of a qubit with continuous geometric quantum state $q(Z)$ this is:
\begin{align*}
H_2\left[ Z\right] & = \lim_{\epsilon \to 0}
  \left( H(Z^\epsilon) + 2\log \epsilon  \right)
  \nonumber \\
& =- \int_{\CP{1}}\!\!\! d\nu_{FS}^Z q(Z) \log q(Z)  \nonumber \\
& = -\frac{1}{2\pi}\int_{0}^1 \!\!\! dp \int_{0}^{2\pi} \!\!\! \!\!\!  d\phi  \,\,q(p,\phi) \log q(p,\phi)
  ~.
\end{align*}

We now discuss two different but related interpretations of $H_{\DD}$. The
first one, of purely information-theoretic nature; the second one, of physical
nature.

\subsubsection*{Information-theoretic interpretation of $H_{\DD}$}

Even in the classical setting, there is no unique definition of entropy for
continuous variables \cite{Rao04,Rao05}. From a resource-theoretic perspective,
one can argue that various definitions address slightly different resources.
Thus, indirectly, their interpretation can be given by identifying appropriate
operational meanings.

In our quantum setting, if $q$ is absolutely continuous, then $\mathfrak{D}= 2(D-1)$
and $H_{\DD}[Z]$ provides the most straightforward definition: the differential 
entropy functional, see Ref. \cite{Cove91a}. This is essentially Shannon's 
functional
Eq. (\ref{eq:DiscreteEntropy})
adapted to apply to a probability density, in which the sum changes into an integral. 
This can be proven directly from its definition in Eq. (\ref{eq:GDQE}), using the 
assumption that $q$ is absolutely continuous. In this case we have $p_{\vec{j},\vec{k}} =
\mu_q(\QJK{j}{k})=q(Z_{\vec{j},\vec{k}})V_{D-1}(\epsilon)$, for some
$Z_{\vec{j},\vec{k}} \in Q(\vec{j},\vec{k})$. Therefore:
\begin{align}
& H_{2(D-1)}\left[ Z\right] \nonumber \\
  & \quad = -\lim_{L \to \infty} \left( \sum_{\vec{j},\vec{k}}p_{\vec{j},\vec{k}} \log p_{\vec{j},\vec{k}} - 2(D-1)\log L \right)\nonumber \\
  & \quad = -\lim_{L \to \infty} \frac{1}{L^{2(D-1)}}\sum_{\vec{j},\vec{k}}q(Z_{\vec{j},\vec{k}}) \log q(Z_{\vec{j},\vec{k}}) \nonumber \\
  & \quad = -\int_{\CP{D-1}} \!\!\!\!\!\! d\nu_{FS} \,\, q(Z) \log q(Z)
  ~.
\label{eq:StSpAverage}
\end{align}

While the integral extends to the whole of $\CP{D-1}$, since $\lim_{x \to 0} x
\log x = 0$ only $q(Z)$'s actual support contributes in a nontrivial way.  As
with classical continuous variables, the information-theoretic interpretation
of $H_{\DD}[Z]$ hinges on the \emph{asymptotic equipartition property} (AEP)
and on the fact that it characterizes the ``size''---probability decay rate---of
the stochastic process' typical set.

In short, the geometric formalism facilitates importing, \emph{mutatis
mutandis}, the tools of analog information theory (continuous variables)
into the quantum domain. This holds since we can use classical measure theory
to discuss the information-theoretic aspects of quantum states.

The price paid is that the arena where this occurs, which usually is an
arbitrary sample space, is a manifold with geometric rules dictated by quantum
physics. However, from the geometric standpoint, there is nothing special or
uniquely challenging about complex projective spaces. Thus, one can appeal to
standard results, simply by providing the correct setup.

We will argue now in more detail that this holds for the independent and
identically distributed (\iid) random variables we consider. While somewhat
restrictive, the AEP for \iid random variables is a fundamental result---one
that lays strong and rigorous foundations for more advanced investigations. For
present purposes, a geometric version of the quantum AEP gives the
information-theoretic interpretation of $H_{\mathfrak{D}}[Z]$.

Results on the classical differential entropy are found in Ref. \cite{Cove91a}.
Here, we provide the proper setup and discuss the results for geometric quantum
states.

First, we examine more closely the \iid~assumption. The projective space of
quantum states of identical systems is not the tensor product of the projective
spaces:
\begin{align*}
\mathcal{P}(\mathcal{H}_D^{\otimes N} ) \neq  \mathcal{P}(\mathcal{H}_D)^{\otimes N}
  ~,
\end{align*}
where $\mathcal{H}_D^{\otimes N}$ is the Hilbert space of $N$ qudits, 
$D=\mathrm{dim} ~\mathcal{H}_D$ and $\mathcal{P}(\mathcal{H}_D)^{\otimes N}$ 
is the manifold of tensor product states of $N$ qudits. This is directly seen 
since $\mathcal{P}(\mathcal{H}_D^{\otimes N} ) \sim \mathbb{C}P^{D^N-1}$ while
$\mathcal{P}(\mathcal{H}_D)^{\otimes N} \sim (\mathbb{C}P^{D-1})^{\otimes N}$.

Second, and the key point, the \iid~assumption guarantees that a geometric
quantum state on $\mathcal{P}(\mathcal{H}_D^{\otimes N} )$ is the product of
$N$ identical geometric quantum states on $\mathcal{P}(\mathcal{H}_D)$. More
precisely, given homogeneous coordinates $Z_{\alpha_1,\ldots,\alpha_N}$ on
$\mathcal{P}(\mathcal{H}_D^{\otimes N} )$, the submanifold of $N$ \iid quantum 
states is described by $N$ homogeneous coordinates $\left\{X_{\alpha_i}
\right\}_{i=1}^N$, with $X_{\alpha_i}$ on the $i-$th element $\PH$, such that
$Z_{\alpha_1,\ldots,\alpha_N} = \prod_{i=1}^N X_{\alpha_i}$. Together with the
\iid~assumption, this implies that $q(Z_{\alpha_1,\ldots,\alpha_N}) =
\prod_{i=1}^N q (X_{\alpha_i})$. Geometrically, then, \iid processes live
on tensor products of the Segre variety embedded in
$\mathcal{P}(\mathcal{H}_D^{\otimes N})$.

In this way, using the tools of classical continuous-variable information
theory, one can easily prove the weak law of large numbers. The details are not
particularly insightful, in that they simply reproduce a particular proof of
the weak law of large numbers, and so are given in App.
\ref{app:QuantumWeakLaw}. In turn, this guarantees that the following geometric
asymptotic equipartition property holds for random quantum variables.

\begin{theorem}[G-AEP for \iid~quantum processes]
Let $Z_1,\ldots,Z_N$ be a sequence of \iid~random quantum variables drawn from
$\CP{D-1}$ according to $q(Z)$, then:
\begin{align*}
H_{\DD}[Z] = - \lim_{N \to \infty} \frac{1}{N} \sum_{k=1}^N \log q(z_k)
  ~.
\end{align*}
\end{theorem}

The limit converges weakly in probability; see the proof in App.
\ref{app:QuantumWeakLaw}. The net result establishes that $H_{\DD}$ is a
well-defined quantum information-theoretic entropy, with clear operational
meaning, directly imported from continuous information theory.

Moreover, it is a tool of practical use as an \iid sampling of the quantum
state space produces an ergodic process. Hence, state-space averages can be
evaluated using sequential time averages and vice versa. Here, we do not dwell
more on this matter. However, we mention that a deeper and more comprehensive
analysis of the use of geometric quantum mechanics to describe quantum
stochastic processes is possible and will be reported elsewhere.

\section{Examples}
\label{sec:Examples}

\begin{figure}
\includegraphics[width=\columnwidth]{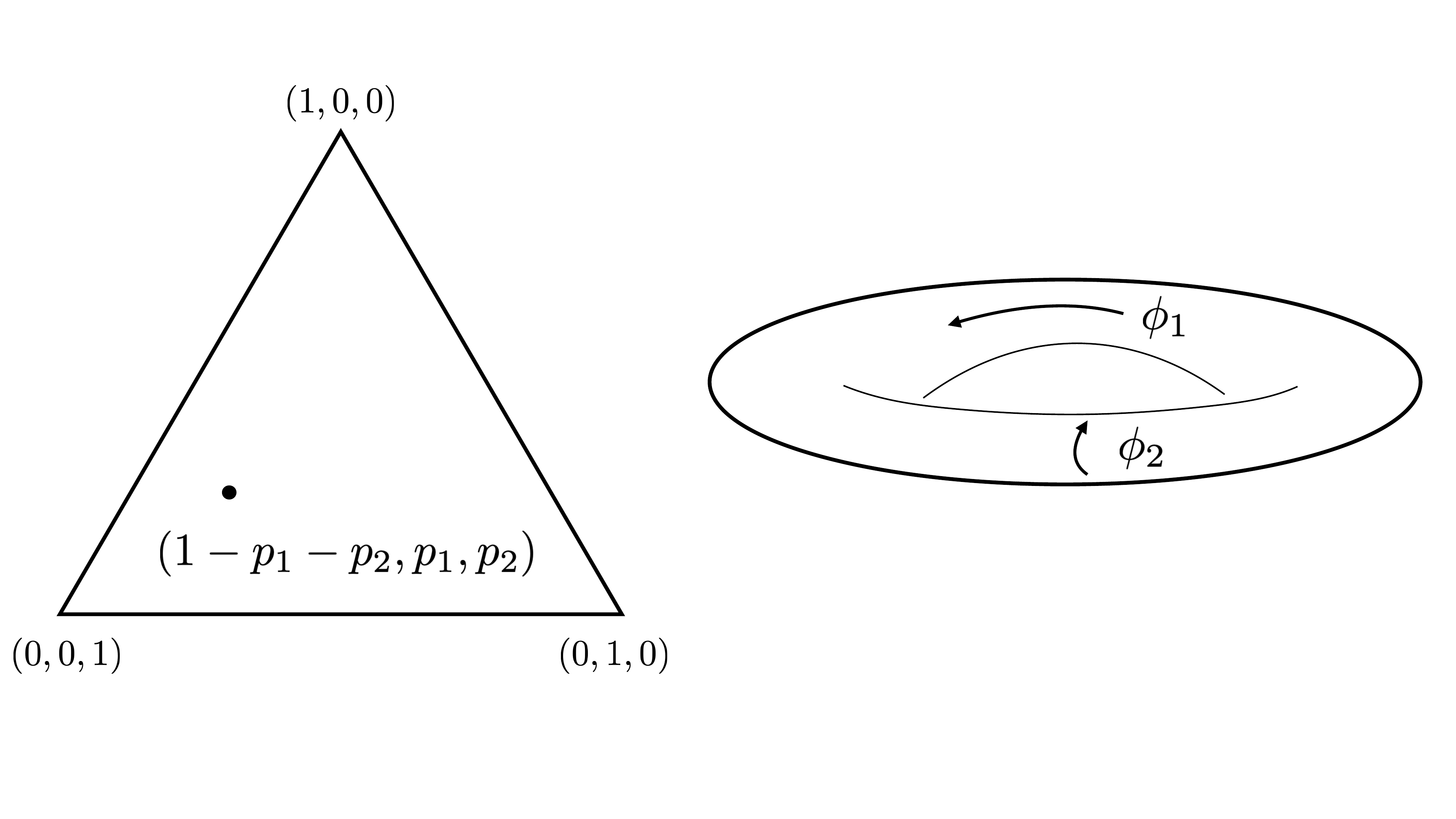}
\caption{Quantum state space of a qutrit: (Left) A finite-dimensional quantum
	system with $D=3$ represented in $2D$. Section \ref{sec:QID} noted that
	canonically conjugated coordinates allow considering the full quantum state
	space as a classical $2-$simplex $\Delta_2$, which represents the space of
	classical probability distributions $(1-p_1-p_2,p_1,p_2)$. (Right) A
	two-torus $\mathbb{T}^2$ that accounts for the nontrivial phases
	$\left(\phi_1,\phi_2\right)$.
	}
\label{fig:QSP} 
\end{figure}

This concludes our technical development of the quantum information dimension
and geometric entropy. The next four subsections show how to compute them in
several concrete physical cases, using a combination of analytical and
numerical techniques:
\begin{itemize}
      \setlength{\topsep}{-2pt}
      \setlength{\itemsep}{-2pt}
      \setlength{\parsep}{-2pt}
\item A quantum system in contact with a finite environment;
\item An electron in a two-dimensional box;
\item Chaotic quantum dynamics and quantum fractals: Baker's and Standard
	maps; and
\item The thermodynamic limit.
\end{itemize} 
Before moving to the actual analysis, to appropriately picture the quantum state space 
Fig. \ref{fig:QSP} gives a visual aid---the representation of the full quantum state space of a
qutrit, i.e., $D=3$.

\subsection{Case 1: Finite Environment}
\label{sec:Example1}

As a first example, consider a system $S$ that is part of a larger system $SE$
of finite dimension. In this setting $S$ develops correlations with a
finite-dimensional environment $E$. Let $d_E$ and $d_S$ denote the dimensions
of the Hilbert spaces $\mathcal{H}_E$ and $\mathcal{H}_S$ of $E$ and $S$,
respectively. Also, assume the overall system $SE$ to be in a pure state
$\ket{\psi} \in \mathcal{H}_S \otimes \mathcal{H}_E$. 

If $\left\{ \ket{a_i}\right\}_{i=1}^{d_S}$ is a basis of $\mathcal{H}_S$ and
$\left\{ \ket{e_\alpha}\right\}_{\alpha=1}^{d_E}$ a basis within
$\mathcal{H}_E$, we can always write \cite{Anza20a}:
\begin{align}
\ket{\psi} & = \sum_{i=1}^{d_S} \sum_{\alpha=1}^{d_E}
  \psi_{i\alpha} \ket{a_i}\ket{e_\alpha} \nonumber \\
  & = \sum_{\alpha=1}^{d_E} \sqrt{p_\alpha^E} \ket{\chi_\alpha^S}\ket{e_\alpha}
~.
\label{eq:rev_Schmidt}
\end{align}

Let $\left\{\Pi^E_\alpha = \mathbb{I}_S \otimes \ket{e_\alpha} \!\!
\bra{e_\alpha}\right\}_{\alpha=1}^{d_E}$ be an arbitrary set of projective
measurements on $E$. Then $p_\alpha^E = \bra{\psi} \Pi^E_\alpha \ket{\psi}$ is
the probability of finding the environment in $\ket{e_\alpha}$. And,
$\ket{\chi_\alpha}$ are the system's post-measurement states, upon finding the
environment in state $\ket{e_\alpha}$.

This implies that we can always write the system's reduced density matrix
$\rho^S \coloneqq \Tr_E \ket{\psi} \!\! \bra{\psi}$ as:
\begin{align}
\rho^S = \sum_{\alpha=1}^{d_E} p_\alpha^E \ket{\chi_\alpha^S}\!\!\bra{\chi_\alpha^S}
  ~.
\label{eq:rho_S}
\end{align}

One can interpret Eq. (\ref{eq:rev_Schmidt}) as a Schmidt-like decomposition in
which the sum runs from $1$ to $d_E$---the dimension of the larger of the two
systems. Note that states $\ket{\chi_\alpha^S}$ do not generally form an
orthogonal set. This environment-induced decomposition of the globally pure
state $\ket{\psi}$ provides a geometric quantum state:
\begin{align}
\mu_S = \sum_{\alpha=1}^{d_E} p_\alpha^E \delta_{{\chi_\alpha}}
  ~,
\label{eq:q_S}
\end{align}
where $\delta_{\chi_\alpha}$ is the Dirac measure with support on
$\chi_\alpha$---the element of $\PH$ corresponding to $\ket{\chi_\alpha}$.

From this we can extract two general results for when a system interacts with
a finite-dimensional, albeit arbitrarily large, environment.

\begin{theorem}
Given a finite-dimensional quantum system $S$ interacting with a
finite-dimensional quantum environment $E$, $S$'s quantum information dimension
$\DD=0$.
\label{theo:2}
\end{theorem}

This is easily seen from Eq. (\ref{eq:q_S}), which is a finite sum of Dirac
measures, thus having support on a finite number of points, which has dimension
zero. This is always true for a system interacting with a finite environment.
We can then draw a general result about the dimensional quantum entropy.

\begin{theorem}\label{theo:3}
Given a finite-dimensional quantum system $S$ interacting with a
finite-dimensional quantum environment $E$, $S$'s dimensional quantum entropy is: 
\begin{align*}
H_{0}\left[ q_S\right] = - \sum_{\alpha=1}^{d_E} p_\alpha^E \log p_\alpha^E
  ~,
\end{align*}
where $p_\alpha^E = \bra{\psi} \Pi^E_\alpha \ket{\psi}$ is the probability of
finding the environment in state $\ket{e_\alpha}$.
\end{theorem}

Two comments are in order.

First, the dimensional quantum entropy is invariant under unitary
transformations operating on the system. This is easily seen as it depends only
on $p_\alpha^E$, which has the required behavior.

Second, $H_{0}\left[ q_S\right]$ in general does (but does not have to) scale with the size 
of the environment:
\begin{align*}
H_{0}\left[ q_S\right] \leq \log d_E
  ~,
\end{align*}
While counterintuitive, this dependence is physically consistent. Indeed, here
we are addressing how the state of a quantum system of size $N_S = \log d_S$
results from its correlations with the state of an environment of size $N_E =
\log d_E$. Since (i) there are $d_E = 2^{N_E}$ distinct environmental states
(say, $\ket{e_\alpha}$) and (ii) via Eqs. (\ref{eq:rev_Schmidt}),
(\ref{eq:rho_S}), and (\ref{eq:q_S}) each specifies a pure state $S$, the
geometric entropy of $S$ scales, at most, with the environment's size.

Moreover, we can also extract a lower bound, provided by $\rho^S$'s von Neumann
entropy. Indeed, among all the geometric quantum states with a given $\rho^S$
there is one corresponding to its spectral decomposition $\rho^S = \sum_j
\lambda_j \ket{\lambda_j}\bra{\lambda_j}$. Therefore:
\begin{align*}
H_0[q]\geq S_{\mathrm{vN}}\left[ \rho^S(q)\right]=\sum_j \lambda_j \log \lambda_j
  ~,
\end{align*}
where we emphasize the dependence of $\rho^S$ on $q$, given by
$[\rho^{S}(q)]_{ij} = \mathbb{E}_q\left[ Z_i \overline{Z}_j\right]$.

The choice of $\ket{e_\alpha}$ reflects physical information about the specific
problem being analyzed. For example, in a thermodynamic setting with
Hamiltonian $H = H_0 + H_{\mathrm{int}}$, with $H_0 = H_S + H_E$, we can choose
$\ket{e_\alpha}$ to be the eigenstates of $H_E$ while $\ket{a_i}$ are the
eigenstates of $H_S$. In this case, if the interaction is weak, the environment
acts as a thermal bath. It settles on a distribution $p_\alpha^E$ quite close to
a thermal equilibrium distribution $p_\alpha^E \propto e^{-\beta e_\alpha}$,
where $e_\alpha$ is the  eigenvalue of $H_E$ corresponding to the eigenvector
$\ket{e_\alpha}$: $H_E \ket{e_\alpha} = e_\alpha \ket{e_\alpha}$. In a
quantum computation, in which the environment performs nontrivial operations on
the system of interest, $\ket{e_\alpha}$ can be chosen to be given by the
computational basis.

\begin{figure*}
\includegraphics[width=\textwidth]{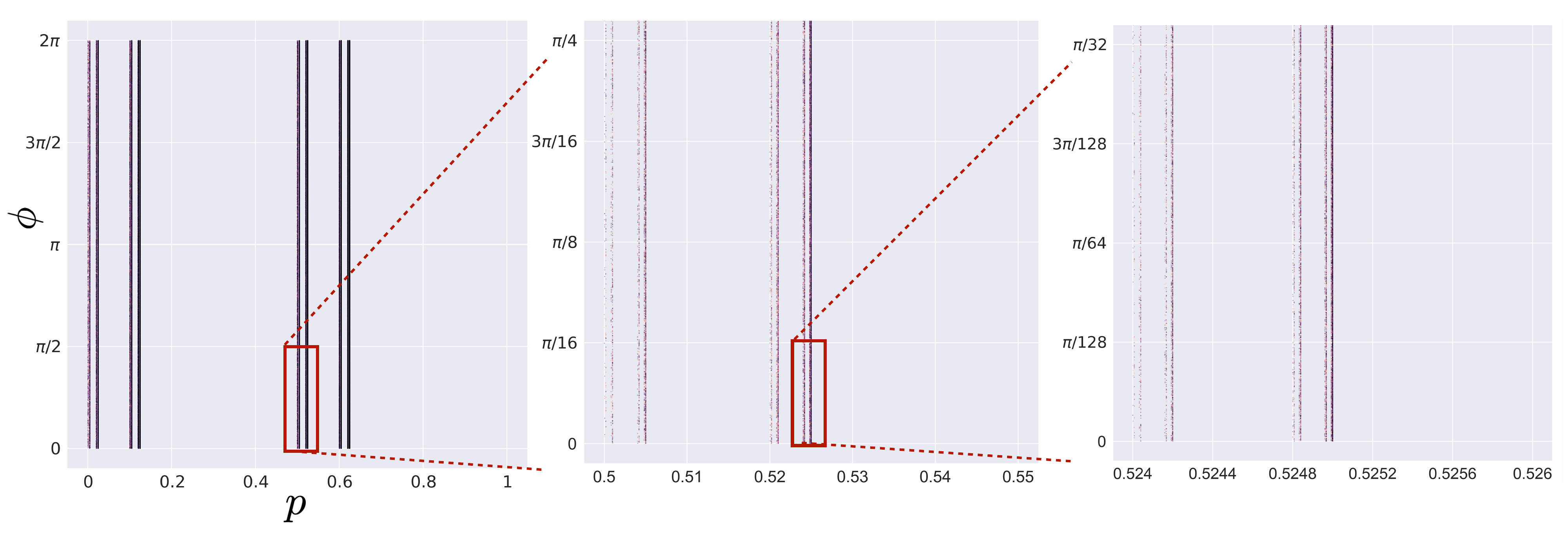}
\caption{Geometric quantum states visited along a single trajectory generated
	by the Extended Baker's Map with parameters $\lambda_a=\lambda_b=0.2$,
	$\beta=4\pi/10$, and initial condition $(p_0,\phi_0)=(0.32865,0.98886)$.
	$N=10^7$ time-steps plotted on the Bloch square $(p,\phi) \in [0,1] \times
	[0,2\pi]$. Over time, due to the map's chaotic nature, even a single
	trajectory covers a (strange) attractor, with self-similar (fractal)
	structure. More specifically, vertically, the attractor has a uniform
	structure. Horizontally, it has self-similar, fractal structure,
	equivalent to a generalized Cantor set. This is demonstrated, going from
	the left panel to the right, via successively magnifying small subsets of
	states.
	}
\label{fig:BM_Attractor} 
\end{figure*}

This first example of calculating the quantum information dimension and
dimensional quantum entropy provides basic intuition about what these
quantities convey about a system's overall behavior resulting from its
correlations with an environment.

\subsection{Case 2: Electron in a 2D Box}
\label{sec:Example2}

Let's now consider a second case in which a finite quantum system interacts
with a quantum system with continuous variables. A concrete example is an
electron confined to move in a $2D$ rectangular box $\mathcal{R}_{2D} =
[x_0,x_1] \times [y_0,y_1]$ where the position and spin degrees of freedom are
assumed to be entangled. The scenario we have in mind is that of an electron
confined to a certain region in which there is a nonhomogeneous magnetic field
$\vec{B}(x,y)$ generating an interaction potential $V(x,y) = (\mu_B g_s /
\hbar) \vec{S} \cdot \vec{B}(x,y)$.

We follow Ref. \cite{Anza20a}'s treatment. Let $\left\{\ket{x,y}\right\}_{x,y}$
be the eigenbasis of the position degrees of freedom and $\left\{
\ket{0},\ket{1}\right\}$ a basis for the spin degree of freedom, Ref.
\cite{Anza20a} showed that a generic state can be written as:
\begin{align*}
\ket{\psi} = \int_{x_0}^{x_1} \!\!\!\!\! dx \int_{y_0}^{y_1} \!\!\!\!\! dy f(x,y) \ket{x,y} \ket{v(x,y)}
  ~,
\end{align*}
with:
\begin{align*}
\int dx dy \vert f(x,y) \vert^2 & = 1 ~, \\
\ket{v(x,y)} & = \sqrt{p_0(x,y)}e^{i\phi_0(x,y)}\ket{0} \\
  & \qquad + \sqrt{p_1(x,y)}e^{i\phi_1(x,y)}\ket{1} ~,\\
p_0 + p_1 & = 1 ~, ~\text{and} \\
(\phi_0,\phi_1)  & \sim (0,\phi_1-\phi_0)
  ~.
\end{align*}
Thus, the spin degree of freedom is described by $f(x,y)$ and $\left\{p_s(x,y),
\phi_s(x,y)\right\}_{s=0,1}$.

The partial trace over the position degrees of freedom, for a generic
$\ket{\psi}$, gives rise to a continuous geometric quantum state, parametrized
by the coordinates $x$ and $y$:
\begin{align*}
\rho^S &= \int dx dy |f(x,y)|^2 \ket{v(x,y)}\bra{v(x,y)}\\
& = \int d\nu_{FS}^{(p,\phi)} q(p,\phi) \ket{p,\phi}\bra{p,\phi}
  ~,
\end{align*}
where $\ket{p,\phi}= \sqrt{1-p}\ket{0}+\sqrt{p}e^{i\phi}\ket{1}$. The second
equality above implicitly defines a distribution on the qubit's projective
Hilbert space. Reference \cite{Anza20a} details the procedure. The following
simply summarizes the final result.

Given an operator $\mathcal{O}$, acting only on the Hilbert space of the spin,
we have the following:
\begin{align}
\MV{\mathcal{O}} & = \int_{x_0}^{x_1} \!\!\! dx \int_{y_0}^{y_1} \!\!\!dy
  |f(x,y)|^2 \mathcal{O}(v(x,y)) \nonumber \\
  & = \int d\nu_{FS}^{(p,\phi)} \, q(p,\phi) \, O(p,\phi)
   ~,
\label{eq:integral}
\end{align}
where $q$ is a geometric quantum state that depends on $f$ and
$d\nu_{FS}^{(p,\phi)} = dp d\phi / 2\pi$ indicates the uniform Fubini-Study
measure with coordinates $(p,\phi)$. The details of how $q$ depends on $f$ and
on the Fubini-Study metric are not immediately relevant, but can be found in
Ref. \cite{Anza20a}. Here, though, we provide an explicit example to illustrate
computing $\mathfrak{D}[q]$ and $H_{\DD}[q]$.

\begin{figure}
\includegraphics[width=\columnwidth]{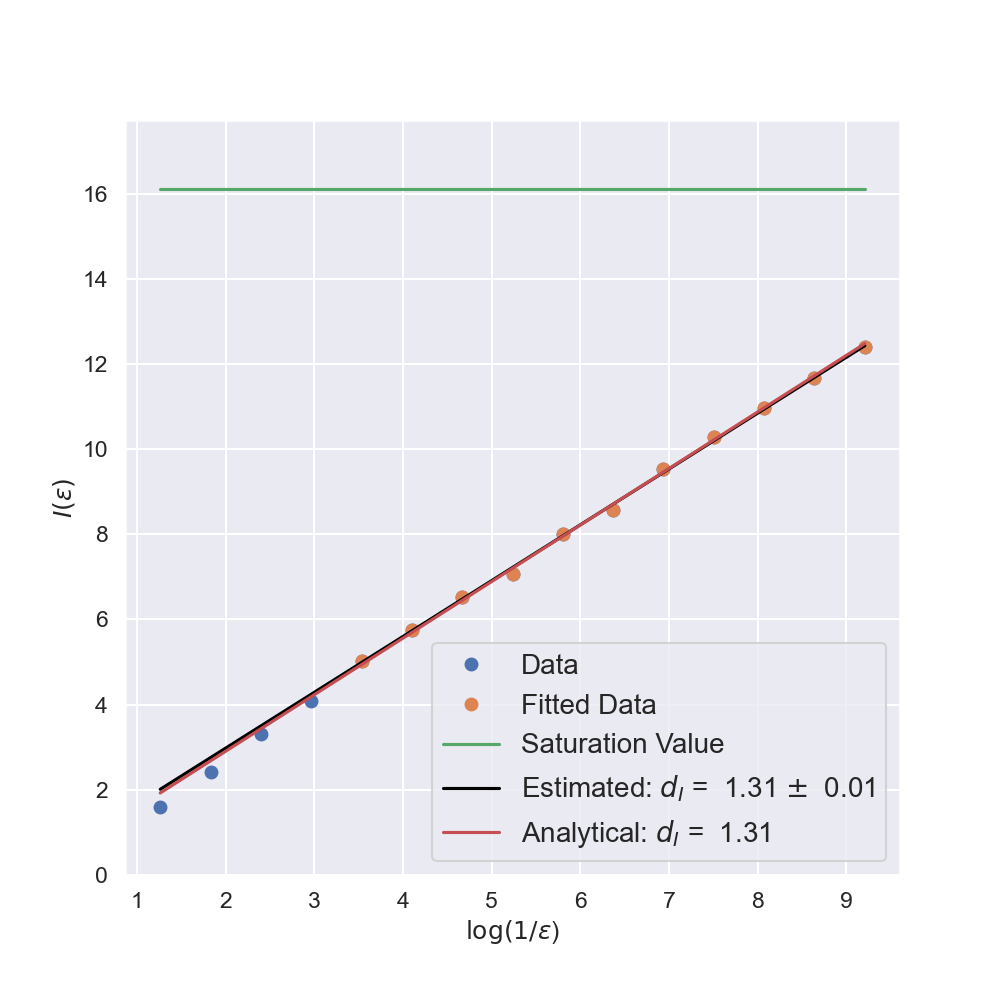}
\caption{Extended Baker's Map information dimension $d_I$: The estimation
	incrementally decreases the coarse-graining scale $\epsilon$ and, at each
	step, calculates $H(Z^\epsilon)$. Then, excluding initial points to avoid
	saturation, it performs a least-square fit to extract the $H(Z^\epsilon)$'s
	growth rate as a function of $\log1/\epsilon$. We estimate $d_I = 1.31 \pm
	0.01$. This is fully consistent with the analytical prediction of $d_I =
	1.31$, plotted in red. See Eq.  (\ref{eq:ID_BM}) and Ref. \cite{Farm83} for
	the analytical estimate.
	}
\label{fig:BM_ID} 
\end{figure}

To be concrete, let $p_1(x,y) = \frac{x-x_0}{x_1-x_0}$, $\phi_1(x,y)
= 2\pi\frac{y-y_0}{y_1-y_0}$, and $f(x,y) = \sqrt{G(x,y)}$, where $G(x,y)$ is a
$2D$ Gaussian on $\mathcal{R}_{2D}$:
\begin{align*}
G(x,y) = \left\{ \begin{array}{ll} 
	\frac{e^{-\frac{1}{2}\left( \frac{x-\mu_x}{\sigma_x}\right)^2}}{\mathcal{N}_x} \frac{e^{-\frac{1}{2}\left( \frac{y-\mu_y}{\sigma_y}\right)^2}}{\mathcal{N}_y}~, & (x,y) \in \mathcal{R}_{2D} \\
	& \\
	0 & \textrm{otherwise}
	\end{array} \right.
  ~,
\end{align*}
where $(\mu_x,\sigma_x)$ and $(\mu_y,\sigma_y)$ are the average and variance
along the $x$ and $y$ axis, respectively. $\mathcal{N}_x$ and $\mathcal{N}_y$
are normalization factors.

\begin{figure*}
\includegraphics[width=\textwidth]{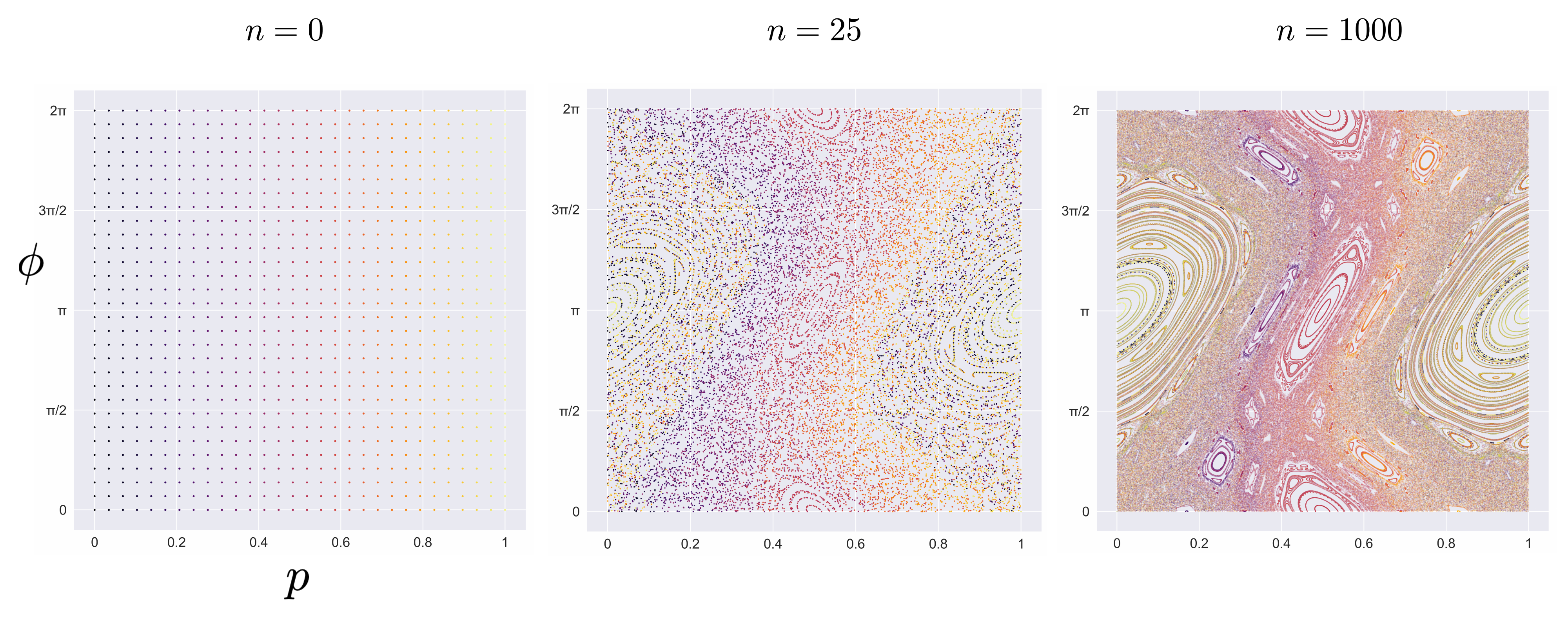}
\caption{Standard Map on the Bloch Square: Quantum states
	$(p,\phi) \in [0,1] \times [0,2\pi]$ iterated at $K=1.15$ over a uniform
	grid of initial conditions. (Left) Initial distribution $n=0$: Homogeneous
	distribution with $30^2$ points $(p_0^{(j)},\phi_0^{(k)}) = (j/30,2\pi
	k/30)$, with $j,k=0,\ldots,29$ running over all initial conditions, each
	distinctly colored. (Middle) After only $n=25$ iterations the points begin
	to mix, according to whether they lie in regions of periodic, quasiperiodic, 
	or chaotic behavior. (Right) The long-term distribution, after $n=1000$ 
	iterations. The full range of behaviors is evident.
	}
\label{fig:SM} 
\end{figure*}

This constructs a geometric quantum state that is absolutely continuous with
respect to $\nu_{FS}$ and therefore expressible via a probability density
$q(p,\phi)$. With the choices made, we obtain:
\begin{align*}
q(p,\phi) = 2 \pi \frac{\mathrm{exp} \left[{-\frac{1}{2} \left( \frac{p - \mu_p}{\sigma_p}\right)^2}\right]}{\mathcal{N}_p}
  \frac{\mathrm{exp}
  \left[{-\frac{1}{2}\left( \frac{\phi - \mu_\phi}{\sigma_\phi}\right)^2}\right]}{\mathcal{N}_\phi} 
  ~,
\end{align*}
with:
\begin{align*}
\mathcal{N}_p \coloneqq \int_0^1 dp
e^{-\frac{1}{2}\left(\frac{p-\mu_p}{\sigma_p}\right)^2}
\end{align*}
and:
\begin{align*}
\mathcal{N}_\phi
\coloneqq \int_0^{2\pi} d\phi
e^{-\frac{1}{2}\left(\frac{\phi-\mu_\phi}{\sigma_\phi}\right)^2}
~.
\end{align*}
Moreover, $\mu_p \coloneqq \frac{\mu_x - x_0}{x_0 - x_1}$, $\sigma_p \coloneqq
\frac{\sigma_x}{x_1 - x_0}$, $\mu_\phi \coloneqq 2\pi \frac{\mu_y - y_0}{y_0 -
y_1}$, and $\sigma_\phi \coloneqq \sigma_y \frac{2\pi}{y_1- y_0}$. 

What are $q(p,\phi)$'s quantum information dimension and the dimensional
geometric entropy? Since this is an absolutely continuous density function,
with support on the whole of $\PH$, one can directly compute the limit in
Eq. (\ref{eq:D}), obtaining $\mathfrak{D} = 2$. Moreover, $H_2[Z]$ assumes a
particularly simple form due to the Gaussian character of $q(p,\phi)$:
\begin{align*}
H_{2}[Z] & = \frac{1}{2}\mathbb{E} \left[ \left(\frac{p-\mu_p}{\sigma_p} \right)^2\right] + \frac{1}{2}\mathbb{E} \left[ \left(\frac{\phi-\mu_\phi}{\sigma_\phi} \right)^2\right] +\\
  & \qquad + \log \mathcal{N}_p + \log \mathcal{N}_{\phi} - \log 2\pi \\
  & = \log \mathcal{N}_p + \log \mathcal{N}_{\phi} + \log \frac{e}{2\pi}
  ~. 
\end{align*}
Thus, again, $\mathfrak{D}=2$ correctly addresses the dimensionality of the
underlying geometric quantum state and $H_2[Z]$ appropriately quantifies its
entropy.

\subsection{Case 3: Chaotic Dynamics and Quantum Fractals}
\label{sec:Example3}

While the examples above clarify the meaning of, and the technology behind,
information dimension, its strength resides in estimating the dimension of
complex probability distributions, especially those whose support is fractal
\cite{Beck93,Dorf99a,Feld12}. These objects have interesting features, such as
structural self-similarity and spontaneous statistical fluctuations, and
they often arise as asymptotic invariant distributions of the dynamics of
complex systems.

The geometric formalism allows us to show how examples imported from the
classical theory of dynamical systems, leading up to fractal invariant sets,
are part and parcel of the phenomenology of quantum systems. In particular, by
exploiting the fact that the Fubini-Study uniform measure on $\mathbb{C}P^1$ in
$(p,\phi)$ coordinates is proportional to the Lebesgue measure on the square
$[0,1] \times [0,2\pi]$, we look at two well-known examples of chaotic
dynamical systems with chaotic attractors with fractal support---the Extended
Baker's Map \cite{Farm83} and Chirikov Standard Map \cite{Chir79}. We show how
to directly implement them in quantum systems by leveraging geometric quantum
mechanics.

\subsubsection{Baker's Map}

First, we look at the Extended Baker's Map (EBM) that, despite the chaotic
behaviors it generates, can be analytically solved. For a detailed discussion
about its properties, especially those related to the information dimension, we
refer to Ref. \cite{Farm83}.

This map is directly implemented on $\mathbb{C}P^1$ via the following unitary
transformations. Let $B$ denote the Extended Baker's Map, each iteration of $B$
maps a quantum state $(p,\phi) \in \mathbb{C}P^1$ to one and only one quantum
state $\left(p^\prime,\phi^\prime\right)$.

\begin{definition}[Extended Baker's Map]
\begin{align*}
\left(p^\prime,\phi^\prime\right) =
  \left\{
  \begin{array}{ll}
 \left( \lambda_a p,2\pi\frac{\phi}{\beta}\right) & \textrm{if $\phi \leq \beta $}\\
 & \\
 \left( \frac{1}{2}+\lambda_b p,2\pi\frac{\phi - \beta}{2\pi-\beta}\right) & \textrm{if $\phi > \beta$}
\end{array} \right.
  ~.
\end{align*}
\end{definition}

Here, we use $\lambda_a \leq \lambda_b \leq \frac{1}{2}$ and $\beta \leq \pi$.
Note that the original extended Baker's Map, as in Ref. \cite{Farm83}, is
defined on the unit square $(x,y)\in [0,1]\times [0,1]$. The above adapts it to
the Bloch square via $(p\to x, \frac{\phi}{2\pi} \to y)$. As a result, $\beta$
is renormalized by a factor $2\pi$ with respect to the one $\alpha$ found in
Ref. \cite{Farm83}: $\alpha \to \beta = 2\pi \alpha$.

Since there is a one-to-one correspondence between points of the Bloch square
and points in $\mathbb{C}P^1$, the action $B[(p,\phi)] =
(p^\prime,\phi^\prime)$ can be implemented on $\mathcal{H}$ as a unitary
transformation between an arbitrary pair of input $(p,\phi)$ and output $(p^\prime,\phi^\prime)$ 
states, as follows.

First, on the qubit Hilbert space, given any $\ket{\psi}$ there is one and only one 
orthogonal state $\ket{\psi^{\perp}}$, up to normalization and phase. 
Thus, a unitary transformation that maps a generic $\ket{\psi}$ onto $\ket{\phi}$ 
can be directly written as $U = \ket{\phi}\bra{\psi}+\ket{\phi^{\perp}}\bra{\psi^{\perp}}$.

Second, embedding of $\mathbb{C}P^1$ with $(p,\phi)$ coordinates onto the qubit
Hilbert space is given by:
\begin{align*}
(p,\phi) \in \mathbb{C}P^1 \to \ket{p,\phi}
  = \sqrt{1-p}\ket{0}+ \sqrt{p}e^{i\phi}\ket{1} \in \mathcal{H}
  ~.
\end{align*}

With this, given a point $(p,\phi)$, the state orthogonal to
$\ket{p,\phi}$ is simply $\ket{1-p,\phi+\pi}$. This means
$\braket{p,\phi}{1-p,\phi+\pi}= 0$ for all $(p,\phi)$. Hence, this
results in the unitary $U := U(B)$ that implements $B$ on the 
Hilbert space: 
\begin{align*}
&U(p^\prime,\phi^\prime;p,\phi) =\\
&\quad \ket{p^\prime,\phi^\prime}\bra{p,\phi}
  + \ket{1-p^\prime,\phi^\prime+\pi)} \bra{1-p,\phi+\pi)}
  ~.
\end{align*}
As a result, iterates of $U$ implement the EBM on $\mathbb{C}P^1$. Calling
$(p_n,\phi_n)$ the state after $n$ iterations and $U_{n+1}$ the unitary
implementing the $n$-th iteration, we have $U_n \neq U_k$ for $k\neq m$. Thus,
while the definition is the same, at each iteration it is represented by a
different unitary operator $U_n=U(p_{n+1},\phi_{n+1};p_n,\phi_n)$. Assuming
each iteration of the map takes a finite amount of time, the appropriate way to
interpret this mapping is that it is a inhomogeneous (in state space) vector
field on $\mathbb{C}P^1$ or, analogously, on $\mathcal{H}$.

Reference \cite{Farm83} gives a detailed discussion of the map's dynamic
properties. Here, we simply recall that, given an arbitrary initial point
$(p_0,\phi_0)$, as a result of the dynamics, the point moves on a subset of the
entire state space. The natural measure, resulting from the dynamics
over infinite time, is a fractal object. More accurately, the attractor has a
uniform distribution over $\phi$ while it has the structure of an extended
Cantor set with respect to $p$. See Fig. \ref{fig:BM_Attractor} for a plot of
$10^7$ map iterates, illustrating the attractor's self-similar (fractal)
structure.

Moreover, its information dimension $d_I$ is known analytically:
\begin{align}
d_I(\alpha,\lambda_a,\lambda_b)  = 1+\frac{-\alpha \log \alpha - (1-\alpha) \log (1-\alpha) }{|\alpha \log \lambda_a + (1-\alpha)\log \lambda_b| }
  ,
\label{eq:ID_BM}
\end{align}
where $\alpha = \beta/2\pi$. This gives a quantum information dimension $\DD
\approx 1.31$ and it allows us to benchmark the algorithmic procedure we use to
numerically compute the information dimension, a necessary reference for cases
in which $\DD$ is not known. To extract the dimensional entropy we look at the
estimated zero-point of the curve $H(Z^\epsilon)$ as a function of $-\log
\epsilon$. The linear fit gives $H_{\DD} \approx 0.25 \pm 0.15$. See Fig.
\ref{fig:BM_ID}.

\begin{figure}[h]
\centering
\includegraphics[width=\columnwidth]{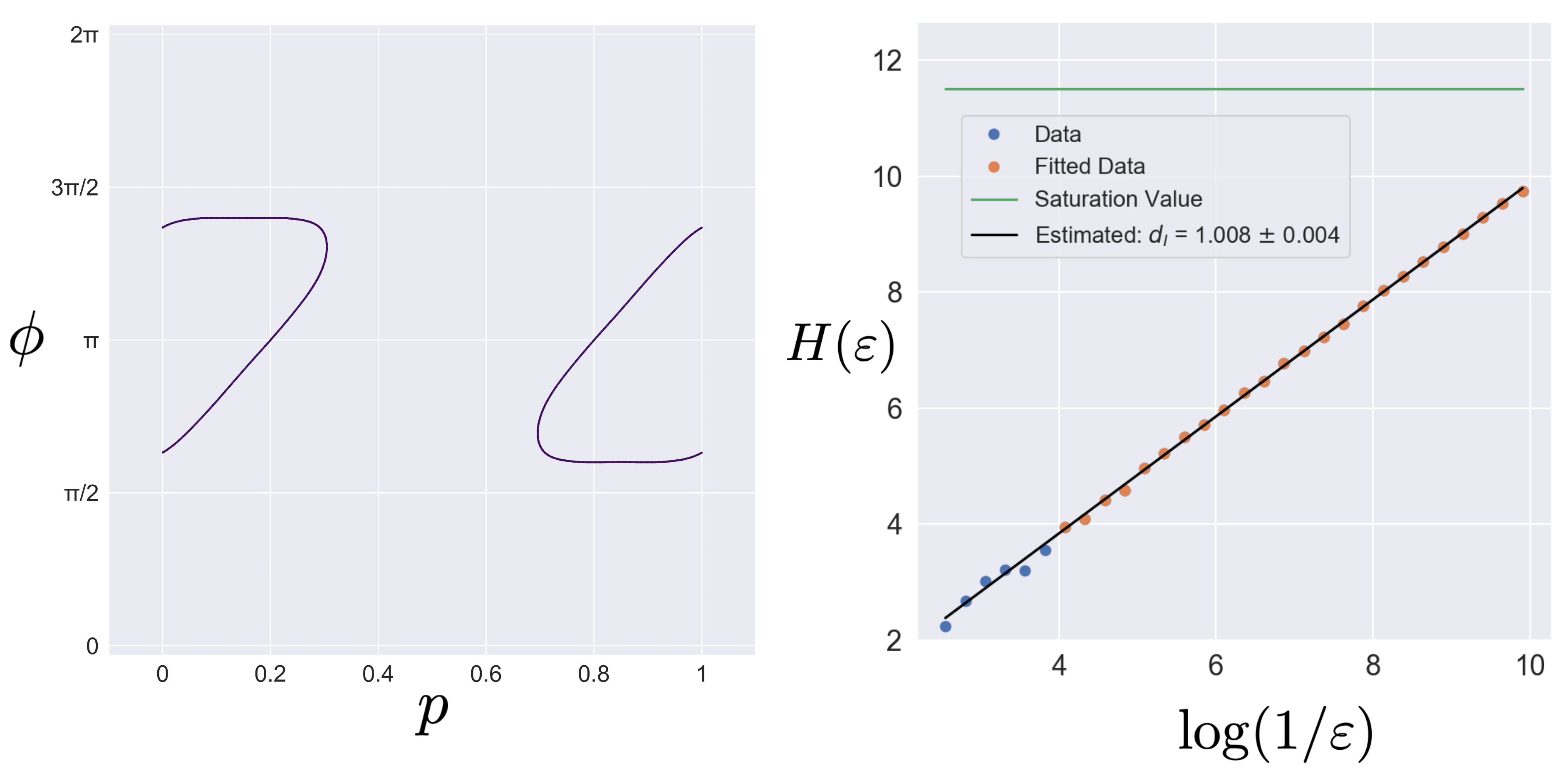}
\caption{(Left) Quasiperiodic orbit with dynamic generated by the Standard Map
	at $K=2$ and initial conditions $(p_0,\phi_0)=(0.2,\pi)$. (Right) Numerical
	estimation of the information dimension, obtained by extracting the growth
	rate of $H(\epsilon)$, shorthand for $H(Z^\epsilon)$, as a function of
	$\log(1/\epsilon)$. The estimated value is consistent, up to $2$
	significant digits, with the expected $\mathfrak{D}=1$. 
	}
\label{fig:ID_SM_Periodic} 
\end{figure}

\begin{figure}[h]
\centering
\includegraphics[width=\columnwidth]{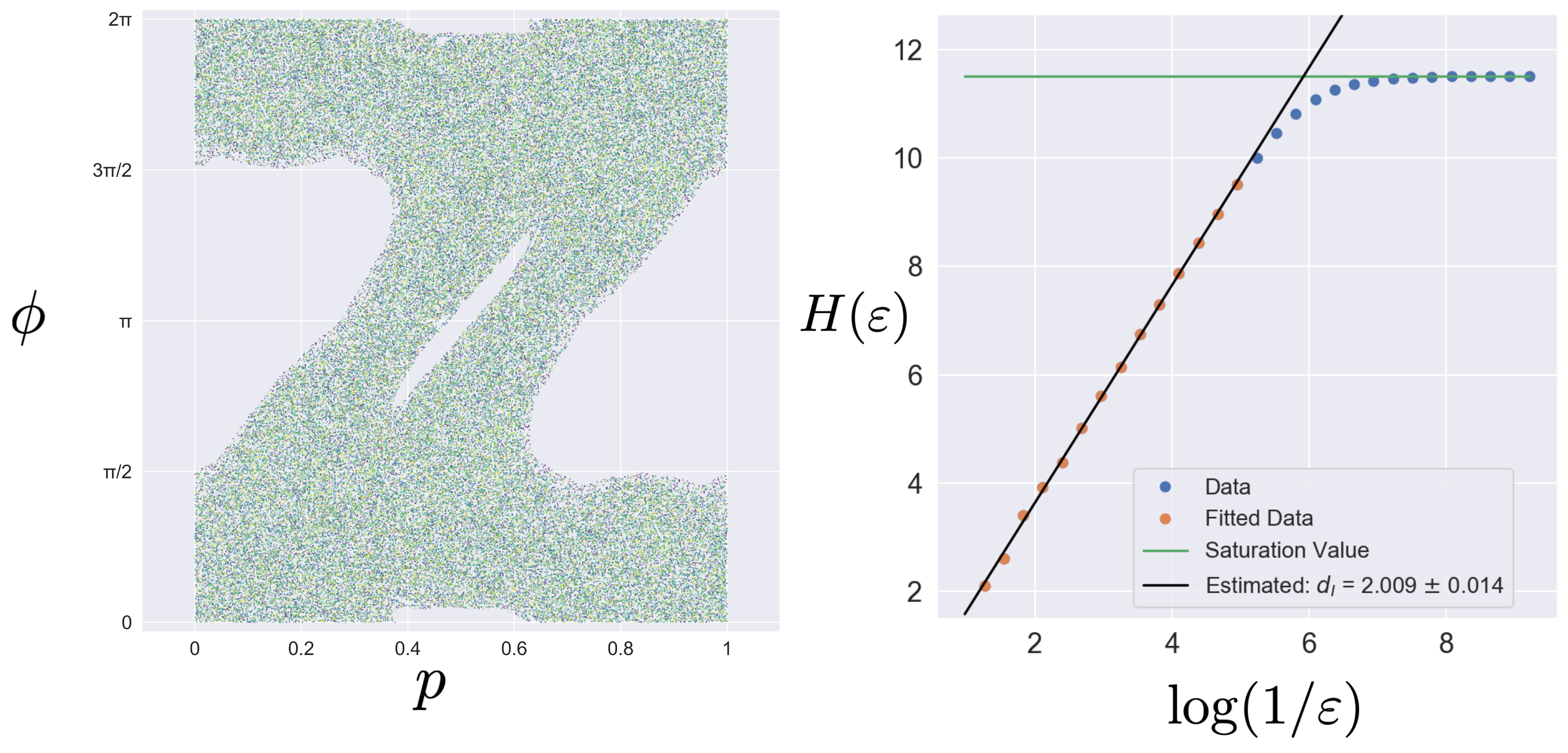}
\caption{(Left) Chaotic orbit with dynamic generated by the Standard Map at
	$K=2$ and initial conditions $(p_0,\phi_0)=(0.1,4\pi/10)$. (Right)
	Numerical estimation of the information dimension, as above. Again, the
	estimated value is consistent, up to $2$ significant digits, with the
	expected one of $\mathfrak{D}=2$. 
	}
\label{fig:ID_SM_Chaotic} 
\end{figure}

\subsection*{Standard Map}

Let's shift attention to the dynamically richer Standard Map (SM). While its
original definition is given on the square of side $2\pi$, it is easily
modified to operate on the Bloch Square $[0,1]\times [0,2\pi]$; i.e.,
$\mathbb{C}P^1$ in $(p,\phi)$ coordinates.

\begin{definition}[Standard Map]
\begin{align*}
\begin{array}{l}
 p^{'} = p + \frac{K}{2\pi} \sin \phi \\
 \\
\phi^{'} = \phi + 2\pi p~,
\end{array}
\end{align*}
\end{definition}
where $p$ is taken modulo $1$, $\phi$ modulo $2\pi$, and $(p_0,\phi_0) \in
[0,1]\times [0,2\pi]$. $K$ is a nonnegative parameter that determines the map's
degree of nonlinearity. Its value is renormalized by $2\pi$ due to the fact that
in its original definition the standard map operates on the unit square $[0,1]\times [0,1]$,
while here we work in $(p,\phi) \in [0,1]\times [0,2\pi]$.

The transformation can be implemented with a set of unitary transformations
$\{S_n\}$, using the same construction just described in Sec. \ref{sec:Example3}
for the EBM.

For $K=0$ only periodic and quasi-periodic orbits are possible. For $K>0$ the
map generates both regions of chaotic behavior and periodic orbits. Increasing $K$,
the extent of periodic orbits decreases, yielding to larger areas of chaotic
behavior. Figure \ref{fig:SM} shows the behavior at $K=1.15$.

As a consequence of the mixed behavior across the state space, the information
dimension of the natural measure, computed over a single trajectory, depends on
the initial condition. If initial conditions lead to chaotic orbits, then we
expect $\mathfrak{D}=2$ while, for periodic orbits, $\mathfrak{D}=1$.

We numerically verify this using the same algorithm exploited in the previous
section to estimate EBM's information dimension. Figures
\ref{fig:ID_SM_Periodic} and \ref{fig:ID_SM_Chaotic} plot the results,
consistent with the expected values.

Analogously, for the dimensional entropy there are two different situations,
depending on whether the initial condition leads to periodic or chaotic
behavior. Since in the chaotic case we simply have a 2D integral, here we
look more closely at the second case, in which $\DD=1$, where the following
treatment can be applied.

\begin{figure}
\includegraphics[width=.4\textwidth]{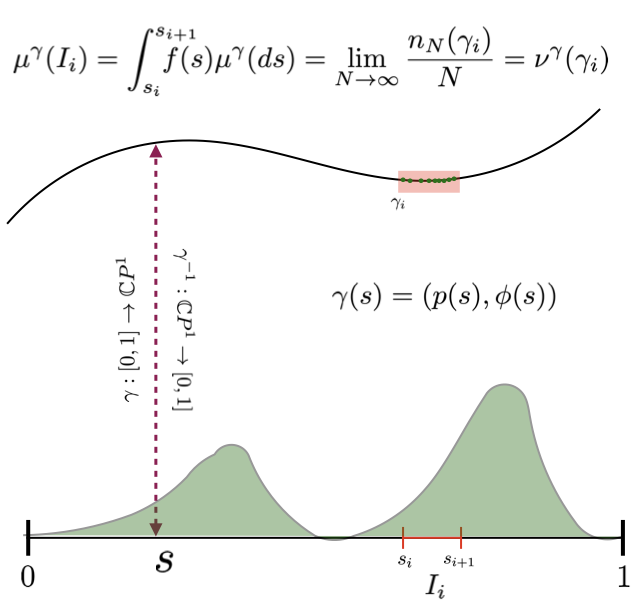}
\caption{Dimensional quantum entropy of a qubit: Geometric construction for
	quasiperiodic behavior. The construction gives rise to a natural measure of
	dimension $\mathfrak{D}=1$. The measure's characteristics on
	$\mathbb{C}P^1$ transfer to a random variable on the unit interval in a way
	that does not deform the distribution: it keeps intact the
	ratios $\nu_\gamma(\gamma_i) / \nu_\gamma(\gamma_j) =
	\mu_\gamma(I_i) / \mu_\gamma(I_j) \approx
	n(\gamma_i) / n(\gamma_j)$. This holds thanks to the fact that
	$\mathfrak{D}=1$ and therefore the support of the measure is a curve
	$\gamma$ on $\mathbb{C}P^1$ that can always be parametrized by
	$\gamma(s)$ with $s\in[0,1]$. The green area represents a fictitious
	probability density on $\gamma$ mapped onto $[0,1]$.
	}
\label{fig:DQE} 
\end{figure}

Referring to Fig. \ref{fig:DQE}, a generic quasiperiodic orbit covers a
$1$-dimensional line, which is identified by a generic equation $f(\phi,p)=0$,
whose solutions are parametrized by a curve $\gamma : [0,1] \to \mathbb{C}P^1$,
or a set of them $\left\{ \gamma_i\right\}$, as in the case of Fig.
\ref{fig:ID_SM_Periodic}. In the following, assume that the set of curves
$\gamma_i$ is bijective, so that given a point $Z$ on any curve, there is one
and only one curve the point is part of, thus the functions $\gamma_i(s)$ admit
inverse $\gamma_i^{-1}: \mathbb{C}P^1 \to [0,1]$. While at first this appears
to be a restrictive assumption, one can always use this construction in cases in
which there are overlapping curves, simply by decomposing them into
nonoverlapping subparts.

Proceed in this way by analyzing each separately and, since the treatment is
formally the same for each, we examine one of them and drop the index $i$.  The
function $\gamma$ is the nonvanishing support of the distribution whose entropy
we are evaluating. On $\gamma$ the proper notion of invariant measure is
provided by the Fubini-Study infinitesimal length element:
$dl_{FS}^{\gamma}\coloneqq ||\dot{\gamma}||_{FS} ds$. Here, $\dot{\gamma} =
(dp/ds,d\phi/ds)$, $|| v ||_{FS} = \sqrt{g_{ab}^{FS}v^a v^b}$ is the
Fubini-Study norm of a vector $v$ in the tangent space, and $g^{FS}$ is the
Fubini-Study metric. Thus, $\gamma$'s Fubini-Study length provides a notion of
measure on $[0,1]$ that is invariant under changes of coordinates and by
unitary transformations in $\mathbb{C}P^1$, via $\mu^\gamma_s(ds) \coloneqq
dl^\gamma_{FS}= ||\dot{\gamma}(s)||_{FS}ds$. This provides the proper notion of
integration on $[0,1]$ to respect all the necessary invariance properties
inherited by the fact that the points on $\gamma$ belong to $\mathbb{C}P^1$.

\begin{figure*}[hbt]
\centering
\includegraphics[width=\textwidth]{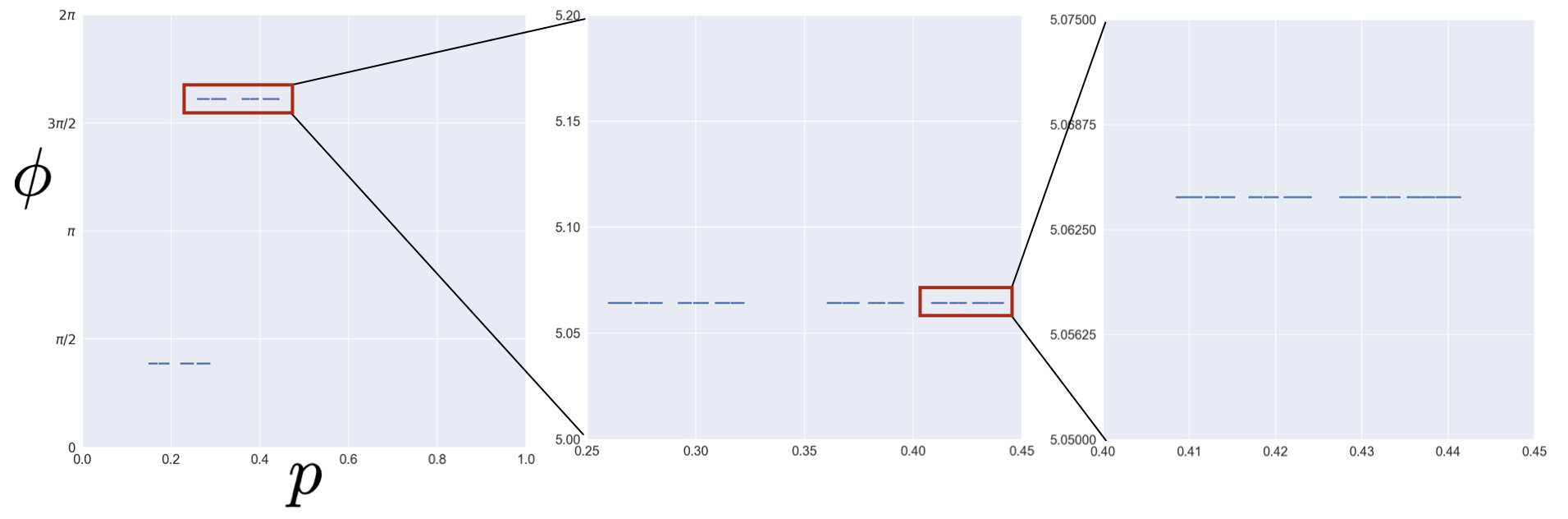}
\caption{Support of the geometric quantum state $\ket{GS(N_E)}$ for the ground
	state of the Heisenberg defect-Hamiltonian with environment size $N=22$.
	The GQS has two separate islands with internal structure that is
	self-similar. (Left to Middle to Right panels) Progressively magnifying the
	region around each part of the support reveals the distribution's
	self-similar support. In the thermodynamic limit its information dimension
	is estimated to be $\mathfrak{D} \approx 0.83 \pm 0.02$.
	}
\label{fig:GQS_HeisenbergDefect} 
\end{figure*}

In this way, given a measure $\nu^\gamma$ on $\mathbb{C}P^1$ with support on
$\gamma$ and density $d\nu^\gamma = \nu(dl^\gamma_{FS}) = f(s) dl^\gamma_{FS}$,
the limit in Eq. (\ref{eq:GDQE}) can be carried out to give:
\begin{align*}
H_1[\nu^\gamma] \coloneqq -\int_0^1 f(s)\log f(s) d\mu_s^\gamma(ds)
  ~,
\end{align*}
where $f(s)$ is the density or, more appropriately, the Radon-Nikodym
derivative, of $\nu^\gamma$ with respect to $\mu^\gamma$.

For example, one can verify that this procedure gives the expected results in
the case of a uniform distribution. Calling $L[\gamma]$ the Fubini-Study length
of curve $\gamma$, we have that $f_{unif}(s) = 1 / L[\gamma]$ and entropy
$\log L[\gamma]$.

It is worth noting that a most important property of this procedure is that it
facilitates computing the entropy of a 1D distribution on $\gamma \in
\mathbb{C}P^1$ by mapping it to the entropy of a continuous density on
$[0,1]$. This amounts to defining $f$ as the continuous density that satisfies
the following consistency constraint: For any arbitrary finite partitioning
$[0,1]=\cup_{i}I_i$ that generates a partition of $\gamma$ into a set $\gamma =
\cup_i \gamma_i$ of $N$ adjacent curves $\gamma_i = \gamma(I_i)$, the density
$f$ is defined via the following chain of equalities:
\begin{align*}
\nu^\gamma(\gamma_i) & = \lim_{N \to \infty} \frac{n_N(\gamma_i)}{N} \\
  & = \int_{I_i} f(s)\mu^{\gamma}(ds) \\
  & = \mu^\gamma(I_i) 
  ~,
\end{align*}
for any $i$ and where $n_N(\gamma_i)$ is the number of points belonging to
$\gamma_i$ in a finite (size $N$) sample of the density on $\gamma$.

This provides a constructive method to analytically compute $H_1$, provided one
has the form of $\gamma$ and $f(s)$. It also gives a direct way to numerically
estimate $H_1$ via the sampling provided by the dynamics: $n_N(\gamma_i) / N
\approx \nu_\gamma(I_i)$ when $N \gg 1$.

\subsection{Case 4: Thermodynamic Limit}
\label{sec:Example4}

Finally, let's shift to explore dimensions for an overtly physical setting: a
finite quantum system without symmetry that interacts with a finite, but
arbitrarily large, environment. The goal is to infer properties in the
thermodynamic limit. The generic procedure to investigate the thermodynamic
limit in geometric quantum mechanics was established and made explicit in Ref.
\cite{Anza20a}.

\begin{figure}[h]
\centering
\includegraphics[width=\columnwidth]{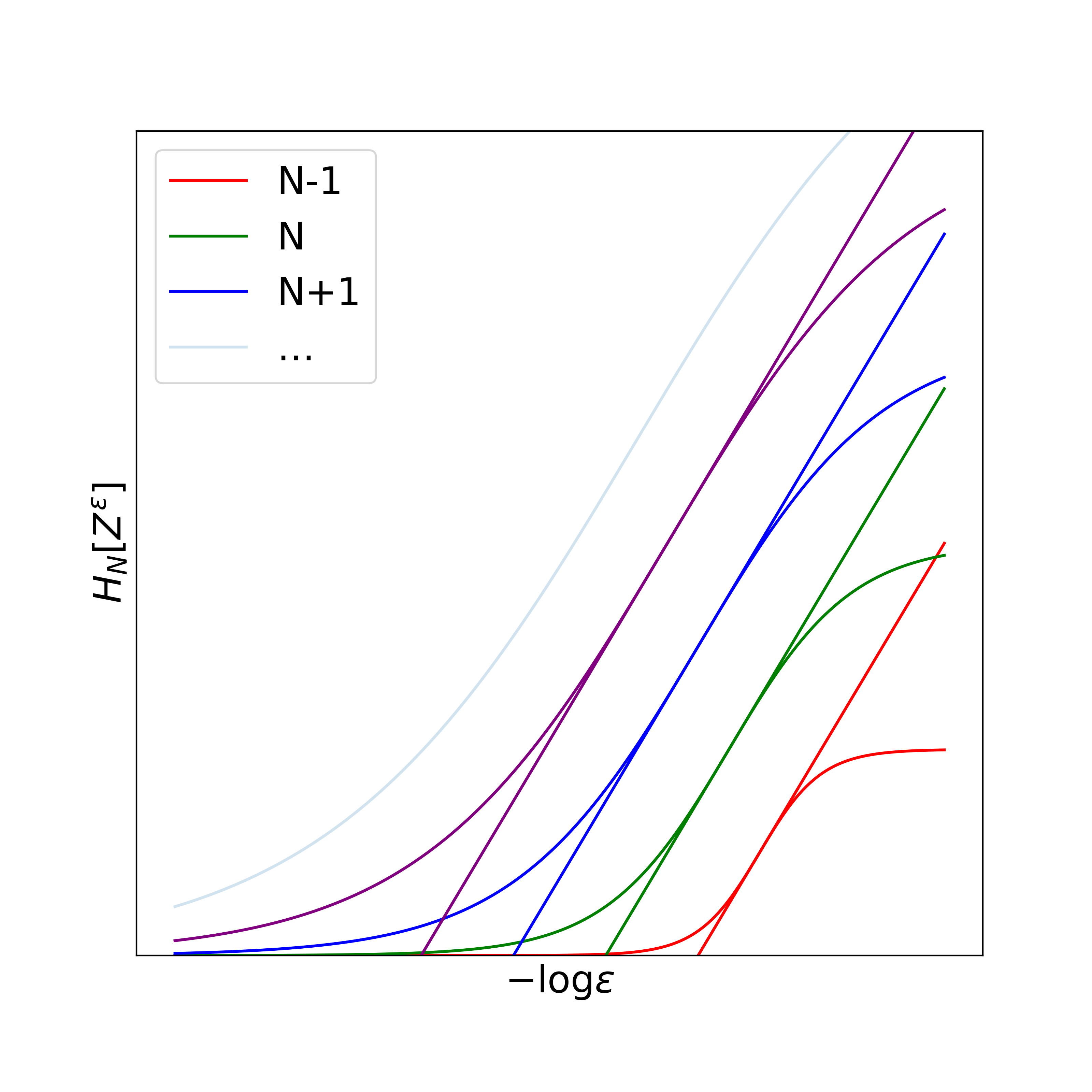}
\caption{Schematic depiction of numerically extracting $\DD_{\infty}$ by
	examining systems with progressively larger system size $N$, while at each
	size we estimate two two scaling curves. At their overlap a linear increase
	with $-\log \epsilon$ is present. This permits estimating $\DD_{\infty}$
	with the slope of the tangent there. In this ideal case, this is exactly
	identical to $\DD_{\infty}$ for each $N$. In reality, finite-size effects
	means the data is noisy and the estimation is harder; cf. Fig.
	\ref{fig:ID_LinearFit}.
	\jpcnote{Explain better why for each $N$ two curves are plotted.}
	\jpcnote{Legend: Add $N+2$ in purple.}
	}
\label{fig:cartoon} 
\end{figure}

The following adopts that procedure and investigates the geometric quantum
state of the ground state of an open-boundary $1D$ spin-$1/2$ Heisenberg chain
with a broken translational symmetry: a defect, realized by removing the local
magnetic field in the last qubit. Let $\vec{\tau}$ be the system's spin
operator and $\vec{\sigma}_j$ the environment's spin operators. The total
Hamiltonian is:
\begin{align*}
H = H_S + H_E(N_E) + H_{\mathrm{int}}(N_E)
  ~,
\end{align*}
where $N_E$ is the size of the environment, $H_S = \vec{B}\cdot \vec{\tau}$, and:
\begin{align}
H_E(N) & =  \sum_{k=1}^{N_E-1} \vec{\sigma}_k \cdot \vec{\sigma_{k+1}}
  + \vec{B} \cdot \vec{\sigma}_k~, \nonumber\\
  H_{\mathrm{int}} & = \vec{\tau} \cdot \vec{\sigma}_1
  ~.
\label{eq:Heis}
\end{align}

The defect-bearing Hamiltonian breaks translational symmetry creating a rich
geometric quantum state---one that exhibits self-similarity and fractal
structure. To illustrate the latter we used $B_z = 0.5$ and $N \in [10,22]$.
As we will see, the choice is supported by the numerical analysis.

At each size $N$ we used the Lanczos algorithm, available in Python via SciPy
\cite{2020SciPy}, to extract the ground state $\ket{\mathrm{GS}(N)}$ and obtain
the associated geometric quantum state $q^{GS}_{N}(Z)$. Figure
\ref{fig:GQS_HeisenbergDefect} plots the support of $q^{GS}_N$ for $N=22$.
Direct inspection of $q_{N}^{GS}(Z)$ suggests that the support of
$q_{\infty}^{GS}$ has a fractal structure with $\DD_{\infty} \in (0,1)$. Thus,
we are interested in $q^{GS}_{\infty}(Z)= \lim_{N\to \infty} q_N^{GS}(Z)$. And
so, for each $N$ we estimate the information dimension via the numerical
procedure used and benchmarked in previous sections. This provides $13$
different datasets to estimate the value of the information dimension in the
thermodynamic limit.

Accurately estimating $\DD_{\infty}$---the QID of $q^{GS}_{\infty}(Z)$---is a
nontrivial since, in principle, it involves evaluating two limits: $\epsilon
\to 0$ and $N \to \infty$. The limits can be singular, meaning that the result
might depend on the order in which they are performed. This is indeed what
happens when trying to directly estimate $\DD_{\infty}$ in a naive fashion.
Namely, by Theorem \ref{theo:2}, at each finite $N$ $\DD_N=0$. This leads one
to conclude that $\DD_{\infty}=0$. This is not correct. The reason a vanishing
dimension appears when first computing $\DD_N$ is that the environment is
finite and, since $H_N[Z^\epsilon]\leq N\log 2$, the curve $H_N(-\log
\epsilon)$ levels off out after the expected linear increase in $-\log
\epsilon$; see Fig. \ref{fig:cartoon}. That is, vanishing dimension arises from
evaluating the limits in the wrong order: $\epsilon \to 0$ at fixed $N$ first
and then $N \to \infty$. Instead, we are interested in the converse:
thermodynamic limit first to obtain $q_{\infty}^{GS}$ and then $\epsilon \to 0$
to extract $\DD_{\infty}$.

Vanishing dimension does not occur when performing the thermodynamic limit
first as this effectively removes the upper bound. If the analytical form of the
ground state in the thermodynamic limit is known, one can proceed without
further ado. This, however, is a rare case and, numerically, these effects are
expected to be present. To cope with this one must correctly identify, for each
$N$, a region of consistent linear growth where $H_N[Z^\epsilon] \approx
-\DD_{\infty} \log \epsilon, \text{~for~all~} \epsilon \in
[\epsilon_{0}(N),\epsilon_1(N)]$. Verifying that the estimate is robust against
increasing the size of the environment yields a reliable estimate of
$\DD_{\infty}$.  A cartoon, of an idealized situation, to provide visual
support to the abstract intuition, is given in Fig. \ref{fig:cartoon}. 

\begin{figure}[h]
\centering
\includegraphics[width=\columnwidth]{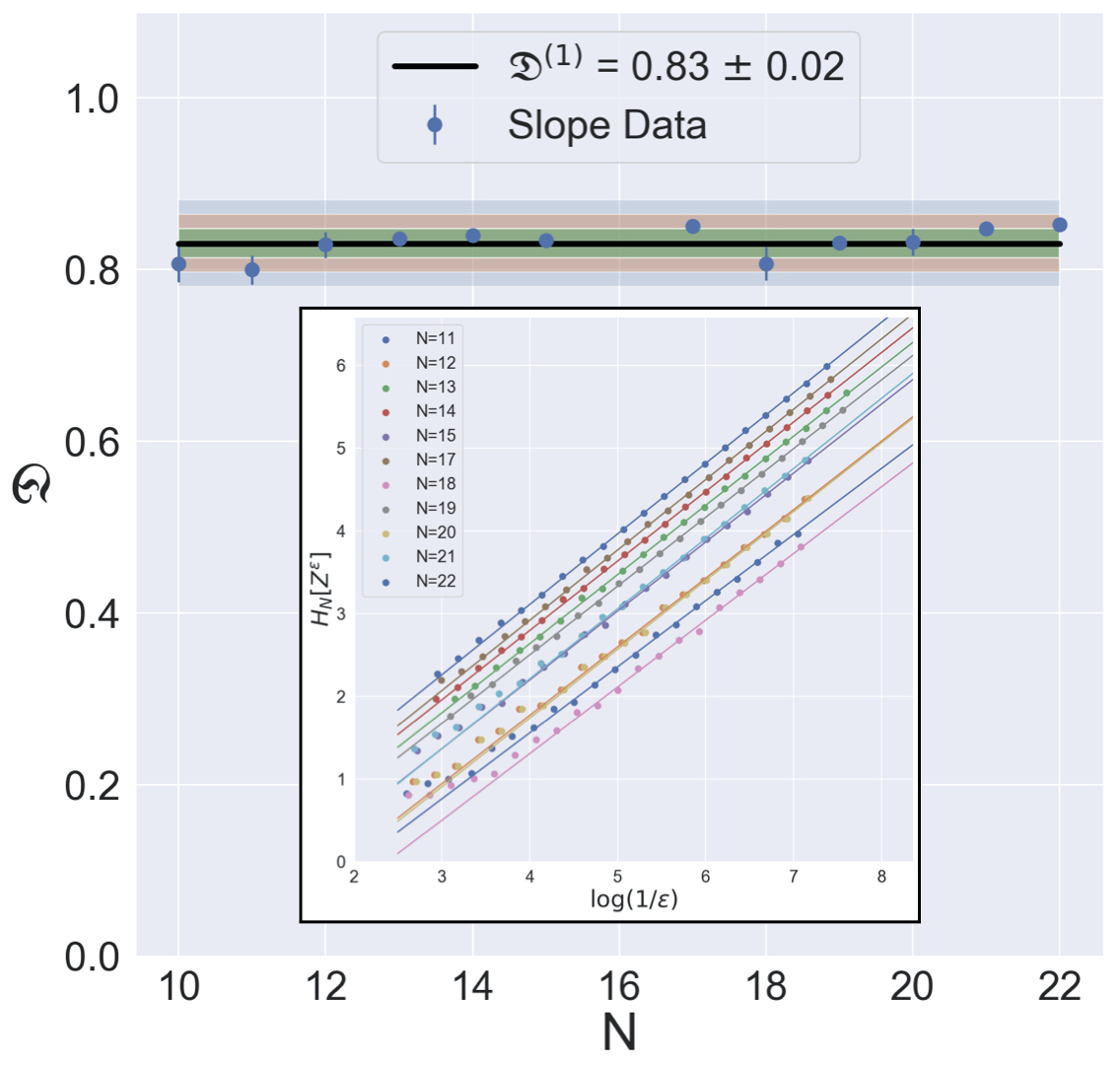}
\caption{Information dimension of the geometric quantum state of a qubit
	interacting with a $1D$ environment with a defective (i.e., nontranslation
	invariant) Heisenberg model of progressively increasing size, see Eq.
	(\ref{eq:Heis}). The entire system is in its ground state $\ket{GS(N)}$,
	where the environment size is $N \in [10,22]$. Each $N$ yields a geometric
	quantum state whose information dimension we estimate using the
	box-counting algorithm explained in the text, extracting the slope using a
	linear fit for $H_N[Z^\epsilon]$. (Inset) A collection of all the data,
	together with the linear fits. (Overall) The collection of horizontal lines
	displays all $M=13$ estimates and extracts the average and standard
	deviation. The result yields $\mathfrak{D}^{(1)}_{\infty}= 0.83 \pm 0.02$. The
	shaded area in green, red, and blue correspond to the areas covered by
	fluctuations around the average of size $\sigma, 2\sigma$, and $3\sigma$,
	respectively, where $\sigma =0.02$ is the standard deviation of the sample
	of slopes.
	}
\label{fig:ID_LinearFit} 
\end{figure}

The estimation was performed by numerically extracting the curves
$H_N[Z^{\epsilon}]$ via a direct box-counting algorithm: fix the value of
$\epsilon$ and build a grid; recall Section \ref{sec:QID}. Then, using the
numerical representation of $q_N^{GS}$, we evaluated the probability mass in
each cell and computed this distribution's Shannon entropy. This gives a
progressively-finer coarse-graining of the state space. The scaling curves were
then analyzed in two separate ways, yielding compatible results.

First, a linear fit was performed by identifying a common region of linearity
for all the $13$ curves $H_N[Z^\epsilon] \propto \mathfrak{D}^{(1)}_N (-\log
\epsilon)$ analyzed. Then, from the $13$ averages we estimated the information
dimension (and its error) from the average and standard deviation.  The results
yield $\mathfrak{D}^{(1)}_{\infty} = 0.83 \pm 0.02$ and are summarized in Fig.
\ref{fig:ID_LinearFit}.

\begin{figure}[h]
\centering
\includegraphics[width=\columnwidth]{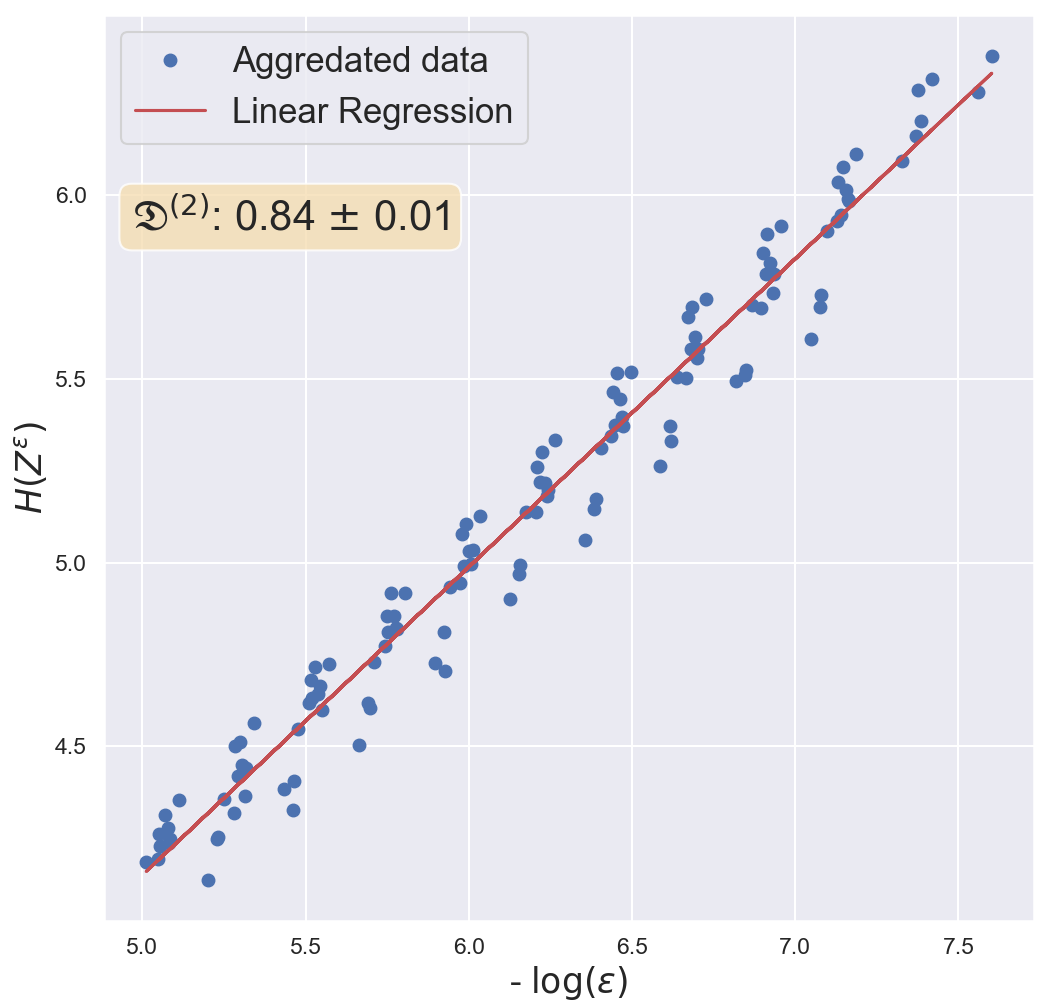}
\caption{Information dimension of the same geometric quantum state, estimated
	using aggregated data. To extract the information dimension in the
	thermodynamic limit, we do not distinguish between points belonging to
	different environment sizes. Aggregating them, we performed linear
	regression to extract a prediction, with associated error, of the
	curve's slope. The result yields $\mathfrak{D}^{(2)}= 0.84 \pm 0.01$.
	}
\label{fig:ID_LinearRegression} 
\end{figure}

Second, we collapsed all the data onto a unique straight line by removing their
estimated vertical offset---setting intercept equal to $0$. We removed a single
outlier, to reduce the error, and checked that this did not appreciably change
the estimate. We then performed linear regression on the aggregated data
points. The result, summarized in Fig. \ref{fig:ID_LinearRegression},
yield $\mathfrak{D}^{(2)}_{\infty} = 0.84 \pm 0.01$.

Altogether, the results support the intuition that the thermodynamic limit is
witness to highly nontrivial geometric quantum states with fractal support.
Increasing the environment's size, the system converges to a self-similar
distribution, with a noninteger information dimension
$\mathfrak{D}_{\infty}\approx 0.83 \pm 0.02$. The state support, shown in Fig.
\ref{fig:GQS_HeisenbergDefect}, is reminiscent of the Cantor set or, more
appropriately, one of its generalizations, e.g., the EBM's invariant
distribution in the $x$ direction.

\begin{figure}[ht]
\centering
\includegraphics[width=\linewidth]{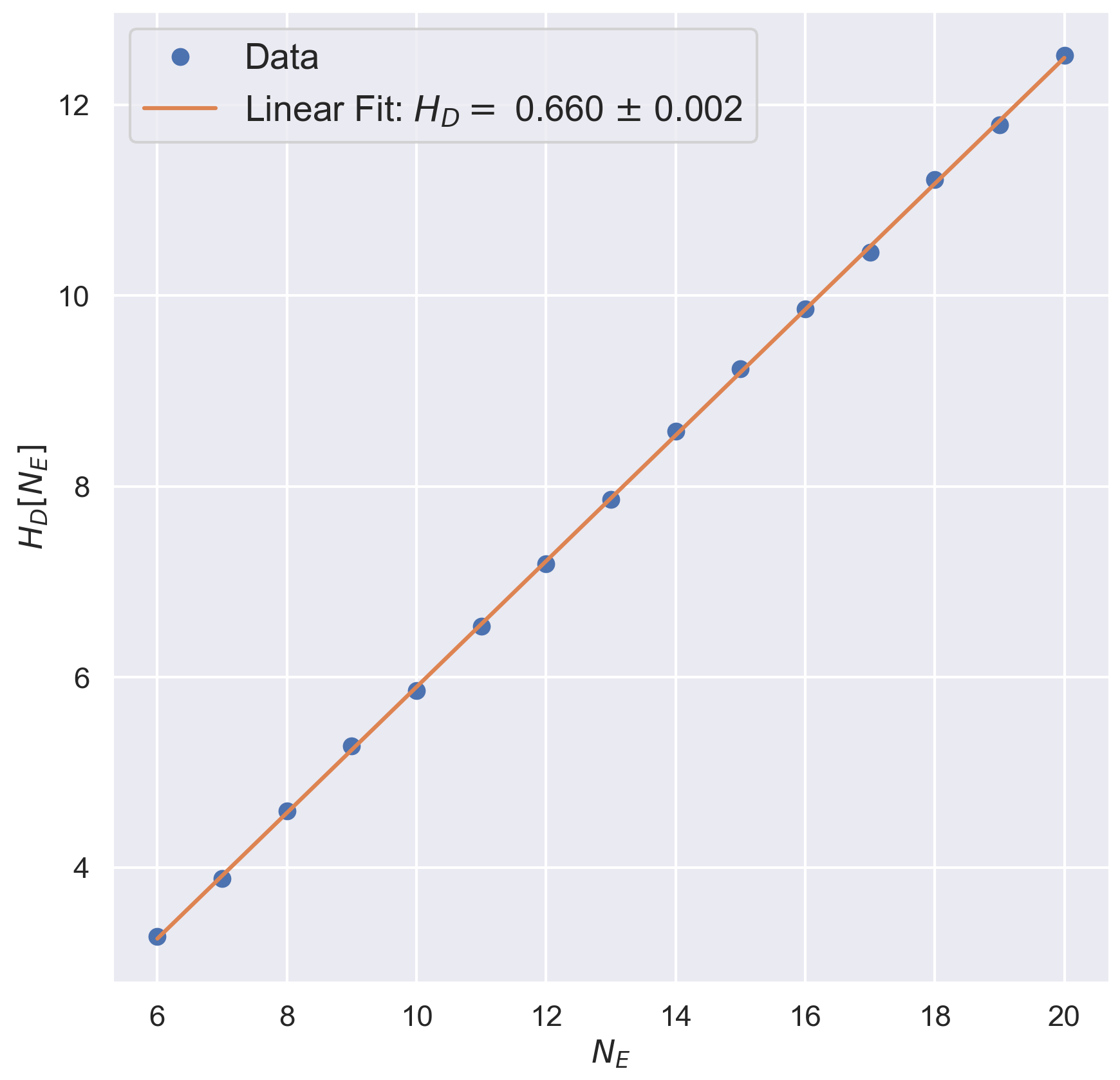}
\caption{Dimensional quantum entropy $H[N_E]$ scaling in the geometric quantum
	state $q$ as a function of environment size $N_E$. The estimated linear
	growth $H[N_E] \sim 0.66 N_E$ and so the state has entropy rate of
	$h_\text{est} = 0.66$.
	}
\label{fig:Entropy_HeisenbergDefect} 
\end{figure}

The estimation of the dimensional geometric entropy is somewhat easier.
Indeed, while its value diverges, it does so in a controlled fashion, which is
at most linear in the environment size $N_E$. We thus extract the entropy rate
by direct inspection of its definition, given explicitly in Thm. \ref{theo:3},
and estimate its linear asymptote, giving the entropy rate $h_\infty$ in the
thermodynamic limit:
\begin{align*}
h_{\infty} & := \lim_{N_E \to \infty} \frac{H(N_E)}{N_E} \\
  & = \lim_{N_E \to \infty} H(N_E)-H(N_E-1)
  ~.
\end{align*}
Since convergence to linear scaling occurred rather rapidly, an accurate
estimate of $h_{\infty}$ is obtained directly from the data, up to two
significant digits: $h_{\infty} \approx h_\text{est} = 0.66 \approx 95\% \log
2$. Figure \ref{fig:Entropy_HeisenbergDefect} gives both the data and the
results of the linear fit.

This concludes our survey of informational properties of geometric quantum
states. Table \ref{tab:summary} summarizes the results. The results leave
several questions and points of discussion, to which we now turn. After which
we draw several conclusions.

\begin{table}[ht]
\small
\begin{tabular}{l|c|c}
& $\mathfrak{D}$ & $H_{\mathfrak{D}}$ \\
\hline
\hline
& & \\
Case 1 (Finite Env)& 0 & $- \sum_{\alpha}p_\alpha \log p_\alpha$  \\
Case 2 ($e^-$ Box)& 2 & $\log \mathcal{N}_p + \log \mathcal{N}_{\phi} + \log \frac{e}{2\pi}$ \\
Case 3 (Baker's Map)  & $1.31 \pm 0.01$ & $0.25 \pm 0.15$  \\
Case 3 (SM-Periodic) & 1 &  $-\int_0^1 f(s) \log f(s) \mu^\gamma(ds)$\\
Case 3 (SM-Chaotic) & 2 & $- \int_{\!\mathbb{C}P^1} \!d\nu_{FS} q(Z) \log q(Z)$\\
Case 4 (Thermo Lim) & 0.83 & $0.66 N_E$\\
& & \\
\hline
\end{tabular}
\caption{Quantum information dimensions and dimensional quantum entropies for
	the geometric quantum states analyzed in Secs.
	\ref{sec:Example1}-\ref{sec:Example4}.
	}
\label{tab:summary}
\end{table}

\section{Discussion}
\label{sec:Discussion}

It has been over a half century since Kolmogorov and followers showed that
Shannon's information theory \cite{Shan48a} provides essential dynamical
invariants for chaotic physical systems
\cite{Gelf56a,Kolm56a,Kolm59b,Kolm58,Kolm59,Sina59}. Today, practically, we
know that information theory readily applies to physical systems that evolve in
discrete time with either a discrete state space or tractable symbolic dynamics
\cite{Crut01a}. Applying the Shannon entropy functional, this involves
quantities that capture informational features with physical relevance---such
as, a system's randomness and structure. This approach has successfully
described the behavior of both Hamiltonian and dissipative classical systems.
That said, the situation is decidedly less straightforward for physical systems
with an inherently continuous sample space that lack a straightforward symbolic
dynamics. Relying on analog information theory, informational descriptions are
markedly more challenging to define and calculate.

Quantum systems belong to this category, as the space $\PH$ of pure states has
continuous nature. Geometric quantum mechanics brings this particular aspect of
quantum systems to the fore, describing their states as probability measures 
on $\PH$---that is, in terms of geometric quantum states. Thus, the geometric
approach directly leads one to adapt the tools from analog information theory
to the quantum domain. 

In this spirit, our development focused on information dimension and
differential entropy, initially proposed by Renyi within the context of analog
information theory. We showed that these tools provide a synthetic view of a
system's geometric quantum state: the information dimension $\mathfrak{D}$
determines the dimensionality of the state's support, while the dimensional
geometric entropy $H_{\mathfrak{D}}$ gives an appropriate differential entropy
for a geometric quantum state with information dimension $\mathfrak{D}$.

Once defined and properly interpreted, we explicitly computed their values in
several examples: a finite-dimensional quantum system interacting with finite
and infinite environments; a qubit evolving with quantum implementations of
nonlinear maps---the Extended Baker's Map and the Standard Map---and finally a
qubit in a progressively larger environment, where we extracted properties in
the thermodynamic limit. 

The interest for these investigations is twofold. On the one hand, extending
the tools of dynamical systems theory to the quantum domain is a topic of broad
and long-lived interest. In point of fact, dynamical systems has led to
successful modeling and quantitative understanding of the structures and
behaviors generated by large classes of synthetic and natural systems---from
nonlinear dynamics to the modeling of population dynamics to tackling the
underlying dynamics of information occurring in a computer running classical
algorithms. On the other hand, the phenomenology of open quantum systems, in
equilibrium and far from equilibrium, is a topic of both fundamental and
applied relevance. Indeed, in the past half decade, the rise of the quantum
computing paradigm made concrete several theoretical investigations focused on
the information-theoretic properties embedded in the dynamics of open quantum
systems and the thermodynamic resources necessary for quantum information
processors to run smoothly and efficiently.

We believe the geometric approach is well suited for these goals, for the
following reasons. The notion of geometric quantum state of a system
\cite{Anza20a} encodes not only the statistics of all measurement outcomes one
can perform on the system, as with the density matrix, but also the detailed
structure of the system-environment quantum correlations that determine said
measurement statistics. Hence, determining the information-theoretic properties
of geometric quantum states gives a novel way to understand the phenomenology
of open quantum systems, whose behavior and structure result from exchanging
information-theoretic and energetic resources with an environment. We believe
this will eventually lead to new analytical tools of power sufficient to
deepen our understanding of the phenomenology of open quantum systems, both in
and out of equilibrium.

\section{Conclusions}
\label{sec:Conclusions}

The development's overtly mathematical nature suggests concluding with three
forward-looking comments.

First, simple examples of geometric quantum states yield an integer value for
$\mathfrak{D}$. At least in the measure theory of classical processes, though,
it is well-known that this is not typical. There are very interesting objects
that exhibit noninteger information dimension: the self-similar or Cantor sets,
now shorthanded as \emph{fractals}. Indeed, these structures are critical to
the operation of Maxwellian demons \cite{Boyd14b} and their modern
realizations---information engines \cite{Boyd15a}. Comparing classical and quantum
domains, it stands to reason that the geometric quantum formalism provides an
interesting arena in which to develop a theory of quantum fractals. Efforts in
this direction are currently ongoing and will be reported elsewhere. The
informational quantities introduced here play a central role in this endeavor.

Second, while here we focused exclusively on $\mathfrak{D}$ and
$H_{\mathfrak{D}}$, it is straightforward to appreciate that the geometric
approach allows for a richer cross-pollination between analog information
theory and quantum information theory. For example, alternative definitions for
core quantities of quantum information theory, based on the geometric approach
and inspired by analog information theory, suggest themselves as parallels of
entropy, relative entropy, mutual information, Kolmogorov-Sinai entropy rate,
excess entropy, bound information, statistical complexity, and many others.
Investigating the relations with their standard quantum counterparts---von
Neumann entropy, quantum mutual information, and the like---presents
interesting challenges. The solutions, we believe, are destined to enrich both
quantum information science and analog information theory.

Third, the geometric approach provides a powerful way to study ensembles of
pure states, with a rich phenomenology to uncover
\cite{Zyck11,Vivo16,Shang15b,Shang15a,Luo11}. This is particularly so given the
recent emergence of quantum information theory and the advances in quantum
computing. These reinforce the need for more advanced tools to study ensembles
of pure states \cite{Cot21,Choi21,WeiHo22}. Indeed, modern quantum simulators
allow extracting ensembles of pure states, like the geometric quantum state, in
systems with a controlled environment. This strengthens the case for the tools
developed here and for the geometric approach more generally.

\section*{Acknowledgments}
\label{sec:acknowledgments}

F.A. thanks Akram Touil, Martin Savage, Sebastian Deffner, Marina Radulaski, Davide
Pastorello, and Davide Girolami for various discussions on the quantum geometric 
formalism. F.A. and J.P.C. thank David Gier, Dhruva Karkada, Samuel
Loomis, and Ariadna Venegas-Li for helpful discussions, Dhruva Karkada for help
with Fig. \ref{fig:GQS}, and the Telluride Science Research Center for its
hospitality during visits. This material is based upon work supported by, or in
part by, Templeton World Charity Foundation Power of Information Fellowship
TWCF0336, FQXi Grant FQXi-RFP-IPW-1902, and U.S. Army Research Laboratory and
the U. S. Army Research Office under grants W911NF-18-1-0028 and
W911NF-21-1-0048.

%
%
\appendix
%
%


\section{Weak law of large numbers for geometric quantum states}
\label{app:QuantumWeakLaw}

The following provides a direct proof of the weak law of large numbers for real
functions on $\PH$. First, let's set up the problem, focusing on quantum
systems with Hilbert space $\mathcal{H}$ of arbitrary dimension $D$. The
projective space of pure states is $\PH$, with dimension $D-1$. A random
quantum variable $Z$ is a triple $(\PH,\mathcal{B},\mu_Z)$ where $\PH$ is the
projective Hilbert space of pure states, $\mathcal{B}$ is its Borel
$\sigma-$algebra and $\mu_Z$ is a measure on $\PH$ such that:
\begin{align*}
\mu(Z) = \int_S \mu_Z(d\nu_{FS})
  ~, 
\end{align*}
where $d\nu_{FS}$ is the normalized Fubini-Study measure:
\begin{align*}
\nu_{FS}(S) & = \int_S d\nu_{FS} \\
  & =\frac{1}{\mathrm{Vol}(\PH)} \int_S dV_{FS} \\
  & =  \frac{\mathrm{Vol}(S)}{\mathrm{Vol}(\PH)}
  ~.
\end{align*}
Now, if $\mu_Z$ is absolutely continuous with respect to $d\nu_{FS}$, we have a
probability density function $d\mu_{Z} = q(z)d\nu_{FS}$.

Let $\{Z_k\}_{k=1}^N$ be a series of $N$ random quantum variables. We are
interested in the case in which the $Z_k$ are all independent and identically
distributed. Take a measurable function $f : \PH \to \mathbb{R}$. Call $X_k
\coloneqq f(Z_k)$. This is a random variable with values on $\mathbb{R}$ with a
law we call $\nu_k$. The $X_k$ with measures $\nu_k$ are also i.i.d.. Thus,
taking $\overline{x}\coloneqq \mathbb{E}[X_k]$ and $\sigma = \mathbb{E}[(X_k -
\overline{x})^2]$, we can define:
\begin{align*}
Y_k & \coloneqq \frac{X_k - \overline{x}}{\sigma}~\text{and} \\
  Y_N & \coloneqq \sum_{k=1}^N \frac{Y_k}{\sqrt{N}}
  ~.
\end{align*}
Now, let $\phi_X(t) \coloneqq \mathbb{E}[e^{itX}]$, we have:
\begin{align*}
\phi_{Y_N}(t) & = \prod_{k=1}^N \phi_{Y_k}(\frac{t}{\sqrt{N}}) \\
  & = \left[\phi_{Y_k}(\frac{t}{\sqrt{N}}) \right]^N
  ~.
\end{align*}

Since we are interested in the limit $N \to \infty$, we expand to see that:
\begin{align*}
\phi_{Y_N}(t) & = \left[\phi_{Y_k}(\frac{t}{\sqrt{N}}) \right]^N \\
  & = \left[1 - \frac{t}{2N} + o\left(\frac{t^2}{N}\right) \right]^N
\end{align*}
means that:
\begin{align*}
\phi_{Y_N}(t) \to \phi_Y(t) \propto e^{-\frac{t^2}{2}}
  ~.
\end{align*}
This is a uniform convergence between characteristic functions that, by means
of the Levy continuity theorem, becomes weak convergence between random variables:
$Y_N \to Y$. Here $Y \sim \mathcal{N}(0,1)$ since the standardized normal
distribution is the only one with characteristic function $\propto
e^{-\frac{t^2}{2}}$. In turn, this means:
\begin{align*}
\overline{X}_N & \coloneqq \frac{1}{N} \sum_{k=1}^N X_k \\
  & = \frac{1}{N} \sum_{k=1}^N f(Z_k)
\end{align*}
implies:
\begin{align*}
\overline{X}_N \sim \mathcal{N}(\overline{x},\sigma_N)
  ~,
\end{align*}
where $\sigma_N \coloneqq \frac{\sigma}{\sqrt{N}}$.

Since $\sigma_N \to 0$, we denote this convergence $\overline{X}_N \to
\overline{x}$, where $\overline{x} = \mathbb{E}[X_k] = \mathbb{E}[f(Z_k)]$ and
fluctuations are of the order $\sigma/ \sqrt{N}$. In other words:
\begin{align*}
\frac{1}{N} \sum_{k=1}^N f(Z_k)
\end{align*}
converges to:
\begin{align*}
\mathbb{E}[f(Z_1)] = \int_{\PH} f(z)q(z)d\nu_{FS}
  ~.
\end{align*}

Thus, for example, one can use $f(Z) = - \log q(Z)$, which is measurable as
long as the geometric entropy is finite. In this way, one finds that:
\begin{align*}
\lim_{N \to \infty} \frac{1}{N}\sum_{k=1}^N -\log q(Z_k)
  & = \mathbb{E}[-\log q(Z)] \\
  & = - \int_{\PH} q(Z) \log q(Z) \\
  & = H_{\mathfrak{D}}[Z]
  ~.
\end{align*}

This establishes the weak law of large numbers for a real measurable function
$f$ on $\PH$ and, in turn, provides a direct proof for the quantum AEP.

\section{Electron in a 2D box}

The following lays out the detailed calculations for the geometric quantum
state in Sec. \ref{sec:Example2}: an electron in a $2D$ rectangular box. The
Hilbert space is given by $\mathcal{P}(\mathcal{H}_x \otimes \mathcal{H}_s)$,
where $\mathcal{H}_x$ is the infinite dimensional Hilbert space of a quantum
particle in a $2D$ box, with basis $\left\{\ket{x,y} \right\}_{x,y}$ and $x \in
[x_0,x_1]$, $y \in [y_0,y_1]$. $\mathcal{H}_s$ is a qubit Hilbert space
describing the spin-1/2 degree of freedom with reference basis
$\left\{\ket{0},\ket{1}\right\}$.
 
Thus, the discrete degrees of freedom of the system---simply spin-$1/2$ in this
case---are described by $f(x,y)$ and $\left\{p_s(x,y),
\phi_s(x,y)\right\}_{s=0,1}$. Then, Eq. (\ref{eq:integral}) becomes:
\begin{align*}
\MV{\mathcal{O}} & = \int_{x_0}^{x_1} \!\!\! dx \int_{y_0}^{y_1} \!\!\!dy
  |f(x,y)|^2 \mathcal{O}(v(x,y)) \\
  & = \frac{1}{2}\int_0^1 \!\!\!dp \int_0^{2\pi}\!\!\!\!\!d\phi \, q(p,\phi) \, O(p,\phi)
  ~,
\end{align*}
where, $p_0(x,y) = 1 - p_1(x,y)$, $p_1(x,y) = p(x,y)$, $\phi_0(x,y) = 0$, and $\phi_1(x,y) = \phi(x,y)$.

To be concrete take $p(x,y) = \frac{x-x_0}{x_1-x_0}$, $\phi(x,y) =
2\pi\frac{y-y_0}{y_1-y_0}$, and $ f(x,y) = \sqrt{G(x,y)}$, where $G(x,y)$ is a
$2D$ Gaussian on $\mathcal{R}_{2D}$:
\begin{align*}
G(x,y) = \left\{ \begin{array}{ll} 
	\frac{e^{-\frac{1}{2}\left( \frac{x-\mu_x}{\sigma_x}\right)^2}}{\mathcal{N}_x} \frac{e^{-\frac{1}{2}\left( \frac{y-\mu_y}{\sigma_y}\right)^2}}{\mathcal{N}_y}~, & (x,y) \in \mathcal{R}_{2D} \\
	& \\
	0 & \textrm{otherwise}
	\end{array} \right.
  ~,
\end{align*}
where $(\mu_x,\sigma_x)$ and $(\mu_y,\sigma_y)$ are the average and variance
along the $x$ and $y$ axis, respectively. $\mathcal{N}_x$ and $\mathcal{N}_y$
are normalization factors.

Using the definitions of the embedding functions, we obtain the following set
of spin vectors, parametrized by $\vec{x} = (x,y)$:
\begin{align*}
\ket{v(x,y)} = \sqrt{\frac{x_1-x}{x_1-x_0}} \ket{0} + \sqrt{\frac{x-x_0}{x_1-x_0}}e^{i2 \pi \frac{y-y_0}{y_1-y_0}}\ket{1}
  ~.
\end{align*}
\begin{widetext}
In turn, this gives:
\begin{align*}
\mathcal{O}(v(x,y)) & = \bra{v(x,y)} \mathcal{O} \ket{v(x,y)} \\
  & = \frac{x_1-x}{x_1-x_0} \mathcal{O}_{00}  + \frac{x-x_0}{x_1-x_0}\mathcal{O}_{11} + \sqrt{\frac{x_1-x}{x_1-x_0}\frac{x-x_0}{x_1-x_0}}  \left(\mathcal{O}_{01} e^{i2\pi \frac{y-y_0}{y_1-y_0}} + \frac{x-x_0}{x_1-x_0} e^{-i2\pi \frac{y-y_0}{y_1-y_0}}\mathcal{O}_{10} \right)
  ~.
\end{align*}

The determinant of the Jacobian matrix between the coordinates $(x,y)$ on $\mathcal{R}_{2D}$ and $(p,\phi) \in [0,1] \times [0,2\pi]$
parametrizing $\mathcal{P}(\mathcal{H}_1^2) \sim \mathbb{C}P^1$ is extracted inverting the functions $p(x,y)$ and $\phi(x,y)$:
\begin{align*}
x(p,\phi) & = x_0 + p(x_1-x_0) ~\text{and}\\
y(p,\phi ) & = y_0 + \frac{\phi}{2\pi}(y_1-y_0)
  ~.
\end{align*}
This gives $D\Phi(Z) = (x_1-x_0)(y_1-y_0) / 2\pi$, which in this case is a
constant. Then, we have, as expected:
\begin{align*}
\mathcal{O}(v(x,y)) & = \mathcal{O}(v(x(p,\phi),y(p,\phi))) \\
  & = (1-p)\mathcal{O}_{00} + p\mathcal{O}_{11} + \sqrt{p(1-p)}\left( \mathcal{O}_{01}e^{i\phi} + \mathcal{O}_{10}e^{-i\phi}\right)
\end{align*}
and 
\begin{align*}
G(x(p,\phi),y(p,\phi)) = \left\{ \begin{array}{ll} 
	\frac{1}{\mathcal{N}_x} \mathrm{exp}\left[{-\frac{1}{2}\left( \frac{x_0 + p(x_1-x_0)-\mu_x}{\sigma_x}\right)^2}\right] \frac{1}{\mathcal{N}_y} \mathrm{exp}\left[{-\frac{1}{2}\left( \frac{y_0 + \frac{\phi}{2\pi}(y_1-y_0)-\mu_y}{\sigma_y}\right)^2}\right]~, & (p,\phi) \in [0,1]\times[0,2\pi] \\
	& \\
	0 & \textrm{otherwise}
	\end{array} \right.
  ~.
\end{align*}
Eventually, using $dV_{FS}^{\mathbb{C}P^1} = \frac{1}{2} dp d\phi$, we can see that $\sqrt{\det g_{FS}(p,\phi)} =
1/2$. Calling $G(x(p,\phi),y(p,\phi)) = \widetilde{G}(p,\phi)$, the geometric
quantum state is:
\begin{align*}
q(p,\phi)
  & = \frac{(x_1 - x_0)(y_1 - y_0)}{2\pi} \times 2 \times \widetilde{G}(p,\phi)
  \\
  & = \frac{(x_1 - x_0)(y_1 - y_0)}{\pi}  \widetilde{G}(p,\phi)
~.
\end{align*}
This can be written as:
\begin{align*}
q(p,\phi) = 2 \frac{1}{\mathcal{N}_p} \mathrm{exp}\left[{-\frac{1}{2}\left( \frac{\mu_p - p}{\sigma_p}\right)^2}\right] \frac{1}{\mathcal{N}_\phi} \mathrm{exp}\left[{-\frac{1}{2}\left( \frac{\phi - \mu_\phi}{\sigma_\phi}\right)^2}\right]
  ~,
\end{align*}
with $(p,\phi) \in [0,1]\times[0,2\pi]$, $\mathcal{N}_p \coloneqq \int_0^1 dp
e^{-\frac{1}{2}\left(\frac{p-\mu_p}{\sigma_p}\right)^2}$, and $\mathcal{N}_\phi
\coloneqq \int_0^{2\pi} d\phi
e^{-\frac{1}{2}\left(\frac{\phi-\mu_\phi}{\sigma_\phi}\right)^2}$. Moreover,
$\mu_p \coloneqq \frac{\mu_x - x_0}{x_0 - x_1}$, $\sigma_p \coloneqq
\frac{\sigma_x}{x_1 - x_0}$, $\mu_\phi \coloneqq 2\pi \frac{\mu_y - y_0}{y_0 -
y_1}$ and $\sigma_\phi \coloneqq \sigma_y \frac{2\pi}{y_1- y_0}$. 

$q(p,\phi)$ is positive and one straightforwardly verifies that it is
normalized. Recall in $(p,\phi)$ coordinates that $dV_{FS}^{(p,\phi)} =
dpd\phi / 2$ and so:
\begin{align*}
\int_{\mathcal{P}(\mathcal{H}_1^2)} \!\!\!\!\!\!\!\!\!\! dV_{FS} \,\, q(p,\phi)
& = 2 \frac{1}{2} \int_{0}^1 dp \frac{1}{\mathcal{N}_p}
e^{-\frac{1}{2}\left(\frac{p-\mu_p}{\sigma_p}\right)^2} \int_0^{2\pi}d\phi
\frac{
e^{-\frac{1}{2}\left(\frac{\phi-\mu_\phi}{\sigma_\phi}\right)^2}}{\mathcal{N}_\phi}
\\
  & = \frac{\mathcal{N}_p}{\mathcal{N}_p}
  \frac{\mathcal{N}_\phi}{\mathcal{N}_\phi} \\
  & = 1
  ~.
\end{align*}
\end{widetext}

\bibliography{chaos}

\begin{thebibliography}{10}

\bibitem{Anza20a}
F.~Anza and J.~P. Crutchfield.
\newblock Beyond density matrices: Geometric quantum states.
\newblock {\em Phys. Rev. A}, 103:062218, 2021.

\bibitem{Sone21}
S.~Deffner A.~Sone.
\newblock {Quantum and classical ergotropy from relative entropies}.
\newblock {\em Entropy}, 23(9), 2021.

\bibitem{Shan48a}
C.~E. Shannon.
\newblock A mathematical theory of communication.
\newblock {\em Bell Sys. Tech. J.}, 27:379--423, 623--656, 1948.

\bibitem{Gelf56a}
I.~M. Gelfand, A.~N. Kolmogorov, and I.~M. Yaglom.
\newblock Towards a general definition of the quantity of information.
\newblock {\em Dokl. Akad. Nauk SSSR}, 111:745--748, 1956.

\bibitem{Kolm56a}
A.~N. Kolmogorov.
\newblock On the {Shannon} theory of information transmission in the case of
  continuous signals.
\newblock {\em IRE Trans. Info. Th.}, 2(4):102--108, 1956.

\bibitem{Kolm59b}
A.~N. Kolmogorov and V.~M. Tikhomirov.
\newblock $\epsilon$-entropy and $\epsilon$-capacity of sets in function
  spaces.
\newblock {\em Uspekhi Mat. Nauk.}, 14:3, 1959.
\newblock (Math. Rev. 22, No. 2890).

\bibitem{Kolm58}
A.~N. Kolmogorov.
\newblock A new metric invariant of transient dynamical systems and
  automorphisms in {Lebesgue} spaces.
\newblock {\em Dokl. Akad. Nauk. SSSR}, 119:861, 1958.
\newblock (Russian) Math. Rev. vol. 21, no. 2035a.

\bibitem{Kolm59}
A.~N. Kolmogorov.
\newblock Entropy per unit time as a metric invariant of automorphisms.
\newblock {\em Dokl. Akad. Nauk. SSSR}, 124:754, 1959.
\newblock (Russian) Math. Rev. vol. 21, no. 2035b.

\bibitem{Sina59}
Ja.~G. Sinai.
\newblock On the notion of entropy of a dynamical system.
\newblock {\em Dokl. Akad. Nauk. SSSR}, 124:768, 1959.

\bibitem{STROCCHI1966}
F.~Strocchi.
\newblock Complex coordinates and quantum mechanics.
\newblock {\em Rev. Mod. Phys.}, 38(1):36--40, 1966.

\bibitem{Miel68}
B.~Mielnik.
\newblock Geometry of quantum states.
\newblock {\em Comm. Math. Phys.}, 9:55--80, 1968.

\bibitem{Kibble1979}
T.~W.~B. Kibble.
\newblock {Geometrization of quantum mechanics}.
\newblock {\em Comm. Math. Physics}, 65(2):189--201, 1979.

\bibitem{Heslot1985}
A.~Heslot.
\newblock {Quantum mechanics as a classical theory}.
\newblock {\em Phys. Rev. D}, 31(6):1341--1348, 1985.

\bibitem{Page87}
D.~Page.
\newblock {Geometrical description of Berry's phase}.
\newblock {\em Phys. Rev. A}, 36, 1987.

\bibitem{And90}
J.~Anandan and Y.~Ahronov.
\newblock {Geometry of quantum evolution}.
\newblock {\em Phys. Rev. Lett.}, 65, 1990.

\bibitem{Gibbons1992}
G.~W. Gibbons.
\newblock {Typical states and density matrices}.
\newblock {\em J. Geom. Physics}, 8(1-4):147--162, 1992.

\bibitem{Ashtekar1995}
A.~Ashtekar and T.~A. Schilling.
\newblock {Geometry of quantum mechanics}.
\newblock In {\em AIP Conference Proceedings}, volume 342, pages 471--478. AIP,
  1995.

\bibitem{Ashtekar1999}
A.~Ashtekar and T.~A. Schilling.
\newblock Geometrical formulation of quantum mechanics.
\newblock In {\em On Einstein's Path}, pages 23--65. Springer New York, New
  York, NY, 1999.

\bibitem{Brody2001}
D.~C. Brody and L.~P. Hughston.
\newblock Geometric quantum mechanics.
\newblock {\em J. Geom. Physics}, 38(1):19--53, 2001.

\bibitem{Carinena2007}
J.~F. Cari{\~{n}}ena, J.~Clemente-Gallardo, and G.~Marmo.
\newblock {Geometrization of quantum mechanics}.
\newblock {\em Theo. Math. Physics}, 152(1):894--903, jul 2007.

\bibitem{Chruscinski2006}
D.~Chru{\'{s}}ci{\'{n}}ski.
\newblock Geometric aspects of quantum mechanics and quantum entanglement.
\newblock {\em J. Physics: Conf. Ser.}, 30:9--16, 2006.

\bibitem{Marmo2010}
G.~Marmo and G.~F. Volkert.
\newblock {Geometrical description of quantum mechanics—transformations and
  dynamics}.
\newblock {\em Physica Scripta}, 82(3):038117, 2010.

\bibitem{Avron2020}
J.~Avron and O.~Kenneth.
\newblock {An elementary introduction to the geometry of quantum states with
  pictures}.
\newblock {\em Rev. Math. Physics}, 32(02):2030001, 2020.

\bibitem{Pastorello2015}
D.~Pastorello.
\newblock {A geometric Hamiltonian description of composite quantum systems and
  quantum entanglement}.
\newblock {\em Intl. J. Geo. Meth. Mod. Physics}, 12(07):1550069, 2015.

\bibitem{Pastorello2015a}
D.~Pastorello.
\newblock {Geometric Hamiltonian formulation of quantum mechanics in complex
  projective spaces}.
\newblock {\em Intl. J. Geom. Meth. Mod. Physics}, 12(08):1560015, 2015.

\bibitem{Pastorello2016}
D.~Pastorello.
\newblock Geometric {Hamiltonian} quantum mechanics and applications.
\newblock {\em Intl. J. Geo. Meth. Mod. Physics}, 13(Supp. 1):1630017, 2016.

\bibitem{Clemente-Gallardo2013}
J.~Clemente-Gallardo and G.~Marmo.
\newblock {The Ehrenfest picture and the geometry of quantum mechanics}.
\newblock {\em Il Nuovo Cimento C}, 3:35--52, 2013.

\bibitem{Feld12}
D.~Feldman.
\newblock {\em Chaos and Fractals}.
\newblock Oxford University Press, 2012.

\bibitem{Bengtsson2017}
I.~Bengtsson and K.~Zyczkowski.
\newblock {\em {Geometry of Quantum States}}.
\newblock Cambridge University Press, Cambridge, 2017.

\bibitem{Renyi59}
A.~Renyi.
\newblock On the dimension and entropy of probability distributions.
\newblock {\em Acta Mathematica Academiae Scientiarum Hungarica}, 10:193--215,
  1959.

\bibitem{Graf00}
S.~Graf and H.~Luschgy.
\newblock {\em Foundations of quantization for probability distributions}.
\newblock Springer, 2000.

\bibitem{Brig19}
A.~le~Brigant and S.~Puechmorel.
\newblock Approximation of densities on {Riemannian} manifolds.
\newblock {\em Entropy}, 21, 2019.

\bibitem{Pag15}
G.~Pages.
\newblock Introduction to vector quantization and its applications for
  numerics.
\newblock {\em ESAIM Proc. Surv.}, 48:29--79, 2015.

\bibitem{Wu10}
X.~Wu and S.~Verdu.
\newblock Renyi information dimension: Fundamental limits of almost lossless
  analog compression.
\newblock {\em IEEE Trans. Info. Th.}, 56:3721--3748, 2010.

\bibitem{Farm83}
J.~D. Farmer, E.~Ott, and J.~A. Yorke.
\newblock The dimension of chaotic attractors.
\newblock {\em Physica}, 7D:153, 1983.

\bibitem{Cove91a}
T.~M. Cover and J.~A. Thomas.
\newblock {\em Elements of Information Theory}.
\newblock Wiley-Interscience, New York, 1991.

\bibitem{Brody2000}
D.~C. Brody and L.~P. Hughston.
\newblock Information content for quantum states.
\newblock {\em J. Math. Physics}, 41(5):2586--2592, 2000.

\bibitem{Brody1998}
D.~C. Brody and L.~P. Hughston.
\newblock The quantum canonical ensemble.
\newblock {\em J. Math. Physics}, 39(12):6502--6508, 1998.

\bibitem{Brody2007}
D.~C. Brody, D.~W. Hook, and L.~P. Hughston.
\newblock {On quantum microcanonical equilibrium}.
\newblock {\em J. Physics Conf. Series}, 67:012025, 2007.

\bibitem{Anza20b}
F.~Anza and J.~P. Crutchfield.
\newblock Geometric quantum thermodynamics.
\newblock {\em arXiv:2008.08683}, 2020.

\bibitem{Anza20c}
F.~Anza and J.~P. Crutchfield.
\newblock Geometric quantum state estimation.
\newblock {\em arXiv:2008.08679}, 2020.

\bibitem{Brody2016}
D.~C. Brody and L.~P Hughston.
\newblock Thermodynamics of quantum heat bath.
\newblock {\em J. Physics A Math. Theo.}, 49(42):425302, 2016.

\bibitem{Rao04}
M.~Rao, Y.~Chen, B.~C. Vemuri, and F.~Wang.
\newblock Cumulative residual entropy: A new measure of information.
\newblock {\em IEEE Trans. Info. Th.}, 50, 2004.

\bibitem{Rao05}
M.~Rao.
\newblock More on a new concept of entropy and information.
\newblock {\em J. Theo. Probability}, 18, 2005.

\bibitem{Beck93}
C.~Beck and F.~Schl{\"o}gl.
\newblock {\em Thermodynamics of Chaotic Systems}.
\newblock Cambridge University Press, 1993.

\bibitem{Dorf99a}
J.~R. Dorfman.
\newblock {\em An Introduction to Chaos in Nonequilibrium Statistical
  Mechanics}.
\newblock Cambridge University Press, Cambridge, United Kingdom, 1999.

\bibitem{Chir79}
B.~V. Chirikov.
\newblock A universal instability of many-dimensional oscillator systems.
\newblock {\em Phys. Rep.}, 52:263, 1979.

\bibitem{2020SciPy}
P.~Virtanen, R.~Gommers, T.~E. Oliphant, M.~Haberland, T.~Reddy, D.~Cournapeau,
  E.~Burovski, P.~Peterson, W.~Weckesser, J.~Bright, S.~J. {van der Walt},
  M.~Brett, J.~Wilson, K.~J. Millman, N.~Mayorov, A.~R.~J. Nelson, E.~Jones,
  R.~Kern, E.~Larson, C.~J. Carey, I.~Polat, Y.~Feng, E.~W. Moore,
  J.~{VanderPlas}, D.~Laxalde, J.~Perktold, R.~Cimrman, I.~Henriksen, E.~A.
  Quintero, C.~R. Harris, A.~M. Archibald, A.~H. Ribeiro, F.~Pedregosa, P.~{van
  Mulbregt}, and {SciPy 1.0 Contributors}.
\newblock {{SciPy} 1.0: Fundamental Algorithms for Scientific Computing in
  Python}.
\newblock {\em Nature Methods}, 17:261--272, 2020.

\bibitem{Crut01a}
J.~P. Crutchfield and D.~P. Feldman.
\newblock Regularities unseen, randomness observed: Levels of entropy
  convergence.
\newblock {\em CHAOS}, 13(1):25--54, 2003.

\bibitem{Boyd14b}
A.~B. Boyd and J.~P. Crutchfield.
\newblock Maxwell demon dynamics: {Deterministic} chaos, the {Szilard} map, and
  the intelligence of thermodynamic systems.
\newblock {\em Phys. Rev. Let.}, 116:190601, 2016.

\bibitem{Boyd15a}
A.~B. Boyd, D.~Mandal, and J.~P. Crutchfield.
\newblock Identifying functional thermodynamics in autonomous {Maxwellian}
  ratchets.
\newblock {\em New J. Physics}, 18:023049, 2016.

\bibitem{Zyck11}
Zyckzkowski et. al.
\newblock {Generating random density matrices}.
\newblock {\em J. Math. Phys. A}, 52, 2011.

\bibitem{Vivo16}
G.~Oshanin P.~Vivo, M.~Pato.
\newblock {Random pure states: Quantifying bipartite entanglement beyond the
  linear statistics}.
\newblock {\em Phys. Rev. E}, 93, 2016.

\bibitem{Shang15b}
J.~Shang, Y.-L. Seah, H.~K. Ng, D.~J. Nott, and B.-G. Englert.
\newblock {Monte Carlo sampling from the quantum state space: II}.
\newblock {\em New J. Phys.}, 17, 2015.

\bibitem{Shang15a}
J.~Shang, Y.-L. Seah, H.~K. Ng, D.~J. Nott, and B.-G. Englert.
\newblock {Monte Carlo sampling from the quantum state space: I}.
\newblock {\em New J. Phys.}, 17, 2015.

\bibitem{Luo11}
S.~Luo, N.~Li, and S.~Fu.
\newblock {Quantumness of quantum ensembles}.
\newblock {\em Theor Math Phys}, 169, 2011.

\bibitem{Cot21}
J.~S. Cotler, D.~K. Mark, H.-Y. Huang, F.~Hernandez, J.~Choi, A.~L. Shaw,
  M.~Endres, and S.~Choi.
\newblock {Emergent quantum state designs from individual many-body
  wavefunctions}.
\newblock {\em arXiv:2103.03536}, 2021.

\bibitem{Choi21}
J.~Choi, A.~L. Shaw, I.~S. Madjarov, X.~Xie, J.~P. Covey, J.~S. Cotler, D.~K.
  Mark, H.-Y. Huang, A.~Kale, H.~Pichler, F.~G. S.~L. Brandao, S.~Choi, and
  M.~Endres.
\newblock {Emergent Randomness and Benchmarking from Many-Body Quantum Chaos}.
\newblock {\em arXiv:2103.03535}, 2022.

\bibitem{WeiHo22}
W.~W. Ho and S.~Choi.
\newblock {Exact Emergent Quantum State Designs from Quantum Chaotic Dynamics}.
\newblock {\em Phys. Rev. Lett.}, 128, 2022.

\end{thebibliography}

\end{document}